\documentclass[]{jfm}

\usepackage{graphicx}
\usepackage{newtxtext}
\usepackage{newtxmath}
\usepackage{natbib}
\usepackage{hyperref}
\usepackage[normalem]{ulem}  
\hypersetup{
    colorlinks = true,
    urlcolor   = blue,
    citecolor  = black,
}

\newcommand{\RomanNumeralCaps}[1]
\linenumbers
\usepackage{subcaption}
\usepackage{caption}
\usepackage{xcolor}

\newcommand{\avg}[1]{\overline{#1}}

\newcommand{\boldvec}[1]{\vec{#1}}
\newcommand{\tbl}[1]{\ensuremath{\text{TBL}_\text{#1}}}
\newcommand{\deltaTBL}{\ensuremath{\delta_\text{TBL}}}
\newcommand{\deltaTKE}{\ensuremath{\delta_\text{TKE}}}
\newcommand{\deltaU}{\ensuremath{\Delta \boldvec{U}}}

\newcommand{\vmin}{\ensuremath{\boldvec{v}_\text{min}}}

\DeclareMathOperator{\sech}{sech}

\title{Effects of mean flow skew on turbulent shear layers. Part I. Numerical investigation}

\author{Vedant Kumar\aff{1},
  Dipendra Gupta\aff{2},
  Gregory Paul Bewley\aff{2}
 \and Johan Larsson\aff{1}\corresp{\email{jola@umd.edu}}}

\affiliation{\aff{1}Department of Mechanical Engineering, University of Maryland, College Park, MD 20742, USA
\aff{2}Sibley School of Mechanical and Aerospace Engineering, Cornell University, Ithaca, NY 14850, USA}

\begin{document}
\maketitle

\begin{abstract}
Skewed turbulent shear layers, formed by the interaction between two non-aligned turbulent boundary layers, are investigated using high-fidelity large eddy simulations in a temporally evolving framework. 
It is argued that a skewed shear layer of this form should be viewed, in the long-time limit, 
in a rotated reference frame
as the superposition of a standard planar shear layer and an orthogonal jet-like component that decays in time.
The skewed shear layer is found to have reduced vertical integral length scale, and the coherent pressure rollers characteristic of shear layers 
undergo transient realignment towards the 
direction orthogonal to mean shear, consistent with the long-time limiting planar shear layer.
Numerical experiments using fictitious test cases indicate that these effects are 
primarily driven through misalignment in the mean flow, and that the two orthogonal flow components in the mean shear frame are only weakly coupled.
\end{abstract}



\section{Introduction}
\label{secIntroduction}
\begin{figure}
  \centering
  \includegraphics[width=0.6\textwidth]{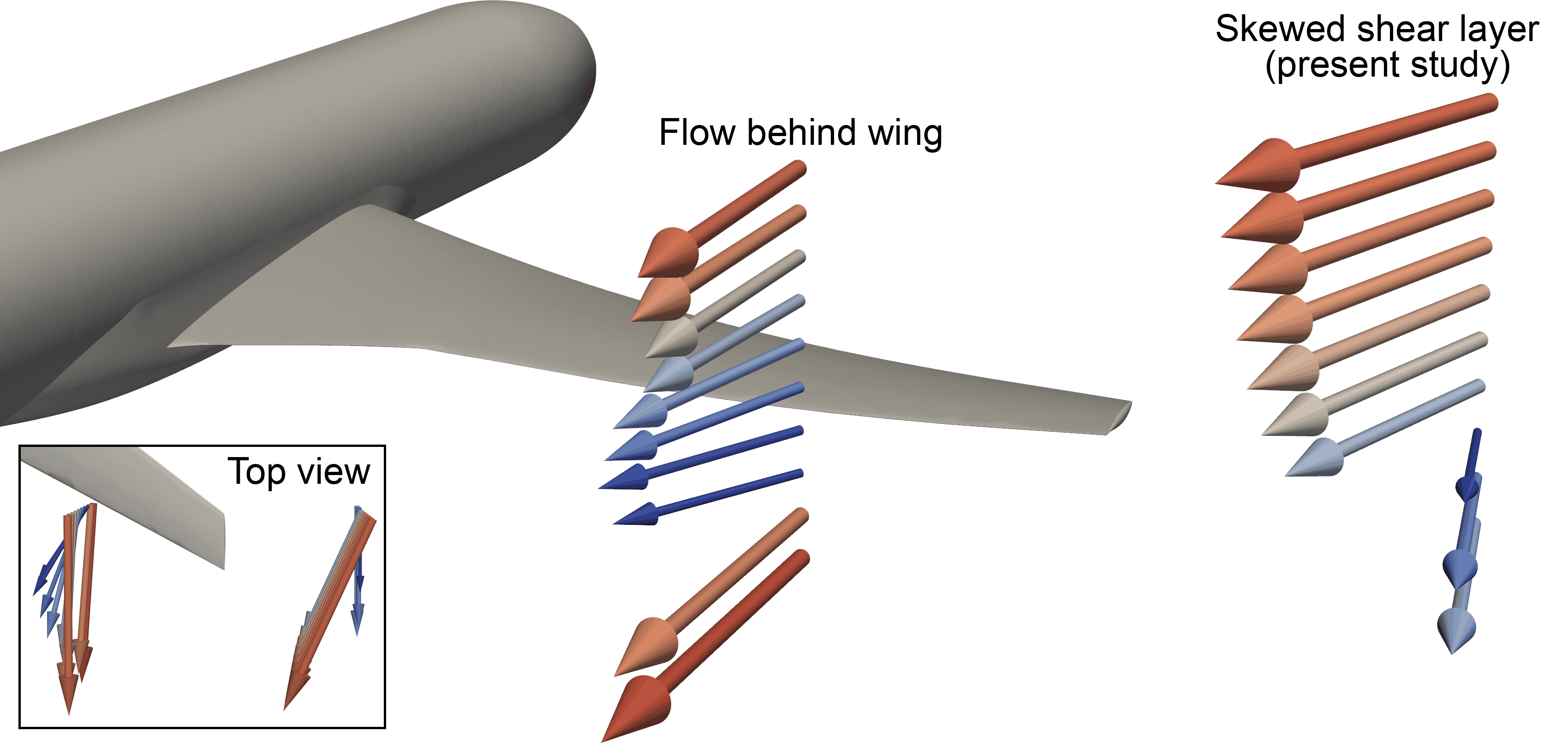}
  \caption{Schematic representation of the mean velocity vectors immediately behind a swept wing (left set of vectors),
  and from the skewed shear layers in the present study (right set of vectors). Arrow colors from blue to red denote increasing 
  velocity magnitude. Inset on the bottom left shows the top view of the same setup.}
\label{figCRMflowVec}
\end{figure}

Turbulent mixing of fluid streams is a common occurrence in geophysical flows, like in the Earth's atmosphere and water bodies, 
as well as in engineering applications.
Shear layers
are one of the fundamental building blocks of turbulent flows, and extensive 
studies over the past decades has informed us of several 
key features such as self-similar growth \citep{Rogers_PoF_1994}, 
the formation of spatially-coherent structures \citep{Brown_JFM_1974}, 
and sensitivity of shear layer growth to initial conditions 
\citep{Bradshaw_JFM_1966,Sandham_JT_2009,Laizet_PoF_2010}, to name a few. 
However, most prior studies have focused on shear layers
where the mean velocity is confined to a plane, i.e., where the mean velocity in the ``spanwise'' direction is zero
(hereafter referred to as planar shear layers).
This simplification, however, 
does not always apply in real-world applications. For example, figure~\ref{figCRMflowVec}
shows the local flow velocity vectors behind a swept aircraft wing,
highlighting that the shear layer that forms between the air from above and below the wing
has finite spanwise velocity with possibly non-aligned freestream vectors
(hereafter, referred to as skewed shear layers).
For comparison, we also overlay the velocity vectors from the flow state generating the skewed shear layers in the present
study in figure~\ref{figCRMflowVec}; the skewed shear layers considered in this study are not meant to model the shear layers behind swept wings or any other specific application, but contain the same qualitative features of a shear layer where the mean velocity vectors are not confined to a plane.

To be more precise, shear layers in real applications are likely three-dimensional with three non-zero components of the mean velocity (3D/3C).
An idealized ``planar shear layer'' is two-dimensional (no variation in the spanwise direction) and two-component (no spanwise mean velocity), thus 2D/2C.
The idealized ``skewed shear layers'' studied here are two-dimensional and three-component, or 2D/3C.

There is limited knowledge on how skewed shear layers differ from their planar counterparts.
In contrast, the effects of having a skewed mean flow have been studied in greater detail for turbulent boundary layers
\citep[see][for an extensive description of relevant studies]{Lozano_JFM_2020}.
Overall, the consensus is that three-component flow in a boundary layer creates 
a misalignment between the Reynolds shear stress and the mean shear vectors \citep{Bradshaw_TRC_1969,Johnston_1970,van_JFM_1975,Bradshaw_JFM_1985}, 
along with a reduction in the Reynolds shear stress for a given amount of turbulence kinetic energy \citep{van_JFM_1975}.
At first, the reduction in shear stress appears counterintuitive since the additional shear 
induced by the spanwise flow should instead lead to increased turbulence production. However, these observations have been 
reconciled with theoretical arguments based on the idea that the near-wall alignment of turbulent eddies in a two-component 
boundary layer is ``optimal'' for Reynolds stress production. A three-component flow distorts this arrangement, 
leading to reduced Reynolds stress \citep{Lohmann_1976_ASME,Bradshaw_JFM_1985,Kiesow_JFM_2003}. Since the directions of a fixed reference 
frame might not be coplanar with the mean shear vector of the three-component flow, \citet{Lozano_JFM_2020}
verified that the reduction in Reynolds shear stress holds true even in a reference frame that aligns with the local mean shear direction. 

There are limited studies focused on three-component effects in shear layers.
Early experimental studies typically generated a skewed shear layer by bringing together
two boundary layers on either side of a splitter plate, with a skew angle between the flows on either side.
For reference, \citet{Hackett_JFM_1970}, \citet{Fric_AIAA_1996}, and, \citet{Azim_EF_2003} considered skew angles of $90^\circ, 39^\circ$, and $18^\circ$, respectively, 
between the two misaligned streams. Overall, these studies agree that skewed shear layers show enhanced mixing rates, especially in the early transient \citep{Fiedler_ETFS_1998}.
Moreover, skewed shear layers also show collapse of mean velocity and Reynolds stress profiles further downstream, similar to planar shear layers.
However, there is a lack of consensus regarding the impact on the Reynolds shear stress, with \citet{Hackett_JFM_1970} and \citet{Azim_EF_2003}
reporting opposite observations of increase and decrease in maximum shear stress for a skewed shear layer, respectively.
Another common method noted in experimental studies has been the use of swirling jets, which induce a skewed shear layer in a curvilinear reference frame.
Notably, \citet{Shiri_AIAA_2008} reported that far downstream, the azimuthal (spanwise) velocity component decays faster than
the streamwise axial velocity component. This results in the azimuthal flow component uncoupling from rest of the flow, 
thus a swirling jet (skewed shear layer) starts behaving like a non-swirling jet (planar shear layer) far downstream,
suggesting that the three-component effects in shear layers are transient in nature.

Apart from lab experiments, \citet{LuLele_JFM_1993} performed a linear stability analysis for an inviscid compressible shear layer, reporting a larger amplification rate for the skewed case, supporting the experimental 
observations of enhanced mixing rates. More recently, \citet{Meldi_JFM_2020} and \citet{Boukharfane_POF_2021} performed 
direct numerical simulations (DNS) for a spatially evolving incompressible and compressible skewed shear layers.
These studies too support the experimental observations of enhanced mixing rates in the early transient of a skewed shear layer. 
However, both were initialized with a hyperbolic tangent mean velocity profile perturbed with random perturbations, thus lacking 
any turbulence structures to begin with.
Since the alignment of turbulence structures potentially explains three-component effects 
in boundary layers and also affect the initial growth of shear layers \citep{Ho_AR_1984,Laizet_PoF_2010}, 
these conclusions could be incomplete.

To summarize, previous studies tend to agree on certain aspects of skewed shear layers, such as enhanced mixing rates in the early 
transient and attaining equilibrium further downstream.
Having said this, we note that several of these prior studies generated skewed shear layers in a way that increased the overall velocity difference between the two freestreams, which would tend to enhance the mixing rates in itself~\citep{Ho_AR_1984}.

The objective of the current study is to develop physical insights into the 
quantitative and qualitative nature of how skew affects turbulent shear layers.
A parallel experimental investigation into skewed shear layers is documented in~\citet{Gupta_TSFP_2024}.

\section{Methodology}\label{secMethod}
\begin{figure}
  \centering
  \includegraphics[width=\textwidth, trim= 0 200 0 110, clip]{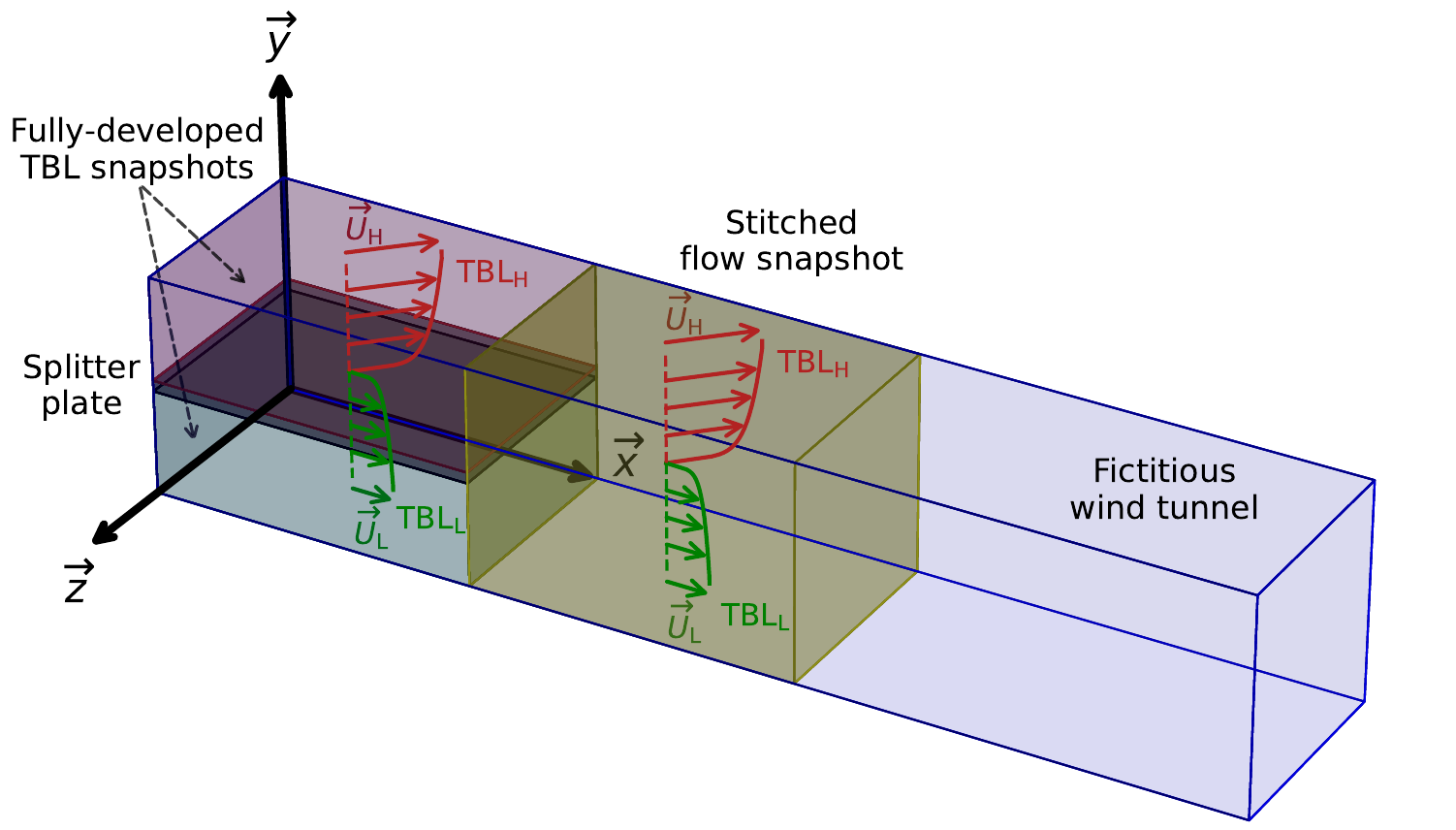}
  \caption{Schematic of the arrangement of computational domains for the 
  temporally evolving boundary layer and shear layer simulations.
  The blue box denotes a fictitious wind tunnel in the fixed lab frame 
  with a splitter plate (colored as gray) on the left boundary.
  The dark-red and green boxes denote the instantaneous snapshots of fully developed 
  turbulent boundary layers $\tbl{H}$ and $\tbl{L}$ on the top and bottom sides 
  of the splitter plate, respectively. These boxes are stacked vertically to generate the 
  initial condition for the shear layer simulation, denoted by the yellow box. 
  }
\label{figSchematic}
\end{figure}

The simulations in this work are performed in a highly idealized way that approximately models the fictitious wind tunnel sketched in figure~\ref{figSchematic}, in which two turbulent boundary layers $\tbl{H}$ and $\tbl{L}$ (for high- and low-speed) develop on either side of a thin splitter plate and come together to form a shear layer that then develops in the $\boldvec{x}$ direction.
Rather than model this directly in the lab frame, the simulations are performed in a frame of reference that moves with constant velocity $\boldvec{U}_\text{c}$, and in which the flow is assumed to be homogeneous (with use of periodic boundary conditions) in the $\boldvec{x}$ and $\boldvec{z}$ directions~\citep{Rogers_PoF_1994,Lozano_JFM_2020}; this is generally referred to as ``temporally evolving'' simulations.
\label{simFramework}
The two boundary layers \tbl{H} and \tbl{L} are thus also solved in a temporally evolving manner, with periodic boundary conditions in both wall-parallel directions.
They are initialized with a laminar Blasius boundary layer solution with added random, dilatation-free, fluctuations.
Once the boundary layers become fully developed and reach a target $99\%$ thickness
$\deltaTBL$, we save their instantaneous snapshots, 
which is represented by the dark-red and green boxes in figure~\ref{figSchematic}.  
The shear layer initial condition is then constructed by vertically stitching the two boundary layer snapshots together, represented by the yellow box in figure~\ref{figSchematic}.
This initial condition at time $t=0$ represents an idealized flow state immediately past the splitter 
plate trailing edge, with an initial shear layer thickness $\delta_\mathrm{0} = 2\deltaTBL$. 
We do note that this approach does not account for the trailing edge wake \citep{Sharma_AIAA_2011},
and therefore does not provide a realistic representation of the flow during the very earliest initial times.
This assumption has been 
used in previous studies \citep{Sandham_JT_2009,Pirozzoli_JFM_2015}, with \citet{Laizet_PoF_2010} showing that
this has minimal impact on the downstream shear layer evolution.
The evolution itself is quantified using the 
shear layer thickness $\deltaTKE$, computed by thresholding the mean turbulence kinetic energy $k = 0.5\avg{u_i^\prime u_i^\prime}$ 
profile on either side of its maxima at the shear layer interface. The threshold is set at $0.0325\max{(k)}$, chosen such that $\deltaTKE/\delta_\mathrm{0} \approx 1$ 
at $t = 0$.
While there are many possible definitions of shear layer thickness, we found that $\deltaTKE$
is more robust than most during the early parts of the evolution.

\subsection{Reference frames}\label{subSecRefFrames}

\begin{figure}
  \centering
  \includegraphics[width=\columnwidth]{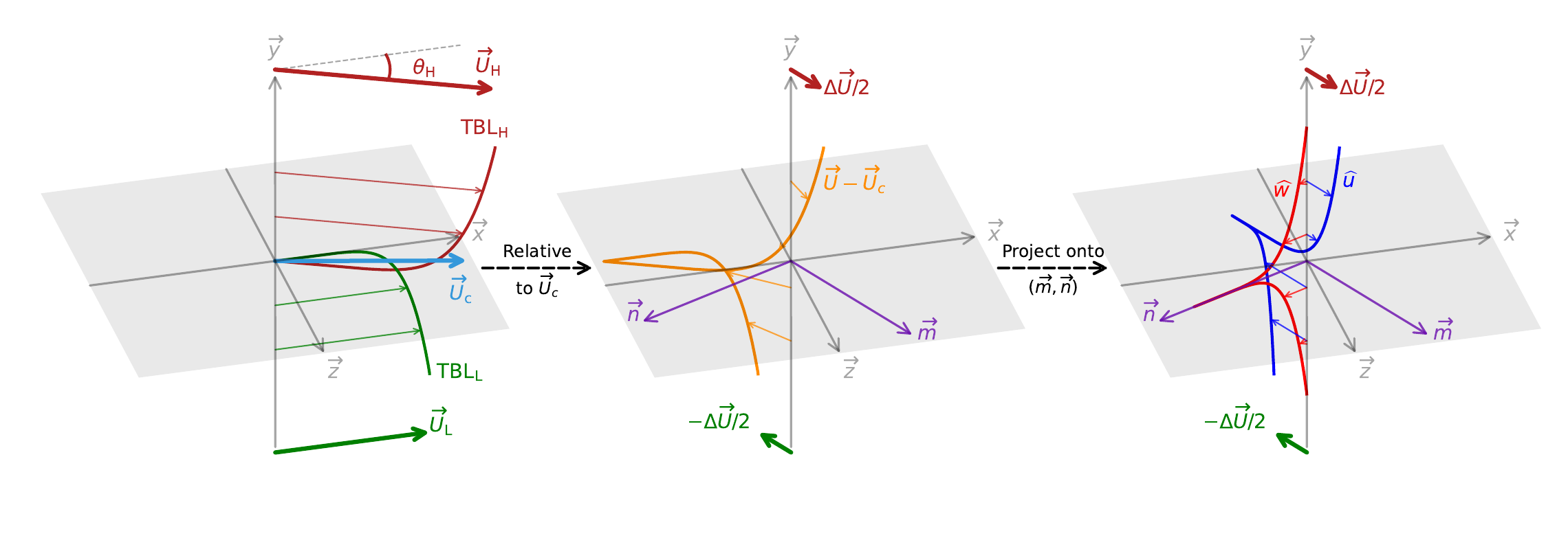}
  \caption{Schematic representation of the transformation from the fixed lab frame $(\boldvec{x},\boldvec{y},\boldvec{z})$ to the 
  convected mean shear frame $(\boldvec{m},\boldvec{y},\boldvec{n})$. The left figure shows representative velocity profiles for \tbl{H} and \tbl{L}
  with freestream velocities $\boldvec{U}_\text{H}$ and $\boldvec{U}_\text{L}$, respectively, in the fixed lab frame. The middle figure shows the 
  flow viewed in a frame moving with the mean convection velocity $\boldvec{U}_\text{c} = (\boldvec{U}_\text{H}+\boldvec{U}_\text{L})/2$,
  resulting in the flow being bounded by equal and opposite freestreams $\pm \Delta \boldvec{U}/2$, 
  where $\Delta \boldvec{U} = \boldvec{U}_\text{H} - \boldvec{U}_\text{L}$. The right figure shows this velocity profile (orange curve
  in the middle figure) decomposed into components $\widehat{u}$ (shear layer-like) and $\widehat{w}$ (jet-like) along directions $\boldvec{m} \parallel \Delta \boldvec{U}$ and $\boldvec{n}\perp\boldvec{m}$, 
  respectively.}
\label{vectorRotationSchematics}
\end{figure}

Similar to previous studies, the flow can be analyzed in a fixed lab frame, defined by the vectors 
$\boldvec{x}$, $\boldvec{y}$ and, $\boldvec{z}$. The early transient evolution
is strongly influenced by the initial condition, making the fixed lab frame 
suitable for its assessment. However, we hypothesize that the long-term evolution of a shear layer should be independent
of the upstream conditions and only depend on the velocity vector difference $\deltaU = \boldvec{U}_\text{H} - \boldvec{U}_\text{L}$.
We therefore define an alternative reference frame aligned with $\deltaU$ using the schematics in figure~\ref{vectorRotationSchematics}.
The left subplot shows representative velocity profiles for \tbl{H} and \tbl{L} on either side of the splitter plate (shaded gray plane)
in the fixed lab frame.

When viewed in a frame moving with the mean convection velocity $\boldvec{U}_\text{c} = (\boldvec{U}_\text{H} + \boldvec{U}_\text{L})/2$ (light blue arrow in left subplot),
the resulting velocity profile $\boldvec{U} - \boldvec{U}_\text{c}$ (orange colored line in the middle subplot) clearly has flow components along 
both the $\boldvec{x}$ and $\boldvec{z}$ directions. In addition, the flow is bounded by equal and opposite freestreams $\pm \deltaU/2$. This motivates defining a pair of 
$\boldvec{m}-\boldvec{n}$ basis vectors (coplanar with the $\boldvec{x}-\boldvec{z}$ plane), such that $\boldvec{m}$ is parallel to $\deltaU$ with 
$\boldvec{n}$ perpendicular to it (both represented using purple arrows in the middle and right subplots). The relative flow can then
be expressed as 
\begin{equation}
  \boldvec{U}-\boldvec{U}_\text{c} = \widehat{u}\boldvec{m} + \widehat{w}\boldvec{n}.
\label{meanShearFrameEq}
\end{equation}
The right subplot shows this decomposition. For a planar shear layer, $\boldvec{m}\parallel\boldvec{U}_\text{H}\parallel\boldvec{U}_\text{L}$, 
hence the flow is entirely along $\boldvec{m}$, represented by the blue profile, and $\widehat{w} = 0$. In contrast, for a skewed shear layer,
$\boldvec{m}\nparallel\boldvec{U}_\text{H}\nparallel\boldvec{U}_\text{L}$, and a finite $\widehat{w}$ component 
is present within the shear layer during the early transient. Outside the shear layer, the freestreams $\pm\deltaU/2$ are parallel to $\boldvec{m}$,
necessitating $\widehat{w} = 0$ in those regions. This results in $\widehat{w}$ having a planar jet-like profile, which is represented by the red profile.
Consistent with our hypothesis that $\deltaU$ dictates the long-term evolution, the shear flow is 
expected to progressively align with $\boldvec{m}$ with downstream evolution, implying that $\widehat{w}$ should decay monotonically.
Consequently, in the long-term evolution limit, a skewed shear layer should become indistinguishable from a planar one 
when viewed in the convected mean shear frame defined by $(\boldvec{m},\widehat{y},\boldvec{n})$. 
Here, $\widehat{y} = y - y_{\widehat{u}=0}$ is the shifted 
transverse coordinate for symmetry between the two freestreams $\widehat{u}/\Delta U = \pm 0.5$,
where $\Delta U = \|\deltaU\|$ is the magnitude of the velocity vector difference. 
Accordingly, we analyze the flow in both the fixed lab frame and the convected mean shear frame 
to assess the different stages of shear layer evolution.
Time will be normalized as $t^* = t \Delta U / \delta_\mathrm{0}$.

\subsection{Flow parameters}\label{secFlowPars}
In the present study, the low-speed stream $\boldvec{U}_\text{L}$ is chosen to be parallel to $\boldvec{x}$ and held fixed for all flow cases, while the high-speed stream $\boldvec{U}_\text{H}$ is adjusted in both magnitude and angle $\theta_\text{H}$ to create different flow cases, with the angle defined in figure~\ref{vectorRotationSchematics}.
The baseline planar case has 
$\boldvec{U}_\text{H} \parallel \boldvec{U}_\text{L} \parallel \boldvec{x}$ (hence, $\theta_\text{H} = 0^\circ$) and 
$\|\boldvec{U}_\text{H}\|/\|\boldvec{U}_\text{L}\|_\mathrm{planar} = 1.5$.
The corresponding boundary layer cases
are labeled
BLU150D00 and BLU100D00, respectively, with the first three digits representing the velocity magnitude 
and the last two digits representing the skew angle (in degrees) of the freestream velocity vector with respect to $\boldvec{x}$.

This introduces three distinct Reynolds numbers, one for each of the boundary layers
(listed in Table~\ref{tableTemporalBLcases})
and one for the initial shear layer state $Re_{\text{TSL},t=0} = \Delta U \delta_\mathrm{0}/\nu = 5000$,
where $\nu$ is the kinematic viscosity of the fluid. 
The design of a skewed shear layer case then presents a design choice, 
as only two of these three Reynolds numbers can be held constant while introducing skew.
Previous studies have typically held the two boundary layer
Reynolds numbers fixed between the planar and skewed shear layers, thereby producing a larger $Re_{\text{TSL},t=0}$ for the latter due to a necessarily larger $\Delta U$ (evident from the vectors in figure~\ref{vectorRotationSchematics}).
In the present study, we instead fix the
shear layer Reynolds number $Re_{\text{TSL},t=0}$ and the low-speed boundary layer Reynolds number,
and accept that the high-speed boundary layer will have different Reynolds numbers in the planar and skewed cases.
Our reasoning is that the shear layer is the central part of this study and that $Re$-effects in boundary layers tend to become smaller for higher Reynolds numbers.
We then choose
$\theta_\text{H} = 25^\circ$ and find 
$(\|\boldvec{U}_\text{H}\|/\|\boldvec{U}_\text{L}\|)_\mathrm{skewed} \approx 1.17$ (labeled BLU117D25)
to maintain a fixed shear layer Reynolds number.

\subsection{Numerical method and grids}\label{secBLsetup}

The simulations are performed using our in-house flow solver \texttt{Hybrid}, which solves the 
compressible Navier-Stokes equations on Cartesian grids using a sixth-order central difference scheme in the split form by 
\citet{Ducros_JCP_2000}.
 The fluid is 
treated as an ideal gas and the subgrid-scale turbulence is modeled using the model by~\citet{Vreman_PoF_2004}. The solver uses the fourth-order 
Runge-Kutta method for explicit time-stepping and a gentle sixth-order artificial dissipation term is added to reduce numerical wiggles 
\citep{Mattsson_JSC_2004}. The solver has been used in many studies for over a decade, with extensive verification and validation efforts~\citep[e.g.][]{Larsson_AIAA_2022}.
Averaging is performed along the homogeneous directions 
with the associated 95\% confidence intervals quantified using the approach of \citet{Oliver_PoF_2014},
shown as shaded regions around selected mean quantities.
The convective Mach number is taken as
$M_c = \Delta U/(c_\text{H} + c_\text{L}) = 0.05$, where $c_\text{H}$ and $c_\text{L}$ are the speeds of sound in the two streams,
which should be sufficiently low to minimize compressibility effects~\citep{Pantano_JFM_2002}.

The length scale for the domain size and the grid-spacing is $\deltaTBL$, the thickness of each boundary layer when they are stitched together to form the initial condition for the shear layer simulation at $t=0$ (the shear layer thickness is then $\delta_\mathrm{0} = 2\deltaTBL$).
The computational domain for the boundary layer simulations is $(L_x, L_y, L_z)/\deltaTBL = (100,100,100)$,
and the domain for the shear layer simulations then becomes $(L_x, L_y, L_z)/\deltaTBL = (100,200,100)$.

The horizontal directions are discretized using 2400 points in both domains, implying $\Delta x = \Delta z = 0.0417\deltaTBL = 0.0208\delta_\mathrm{0}$.
The grid-spacing is approximately twice coarser than in the shear layer DNS of~\citet{Pirozzoli_JFM_2015}.
For the boundary layers, the grid-spacing in viscous units is $\Delta x^+ = \Delta z^+ = 10-14$ as listed in Table~\ref{tableTemporalBLcases}.
The vertical direction is discretized using a geometrically stretched grid with 365 points in each boundary layer (so 730 points in the initial shear layer), with $\Delta y_w^+ = 1.2-1.7$.
Both the boundary layer and shear layer simulations should be viewed as well-resolved large eddy simulations (LES).

\begin{figure}
  \centering
  \includegraphics[width=0.6\textwidth]{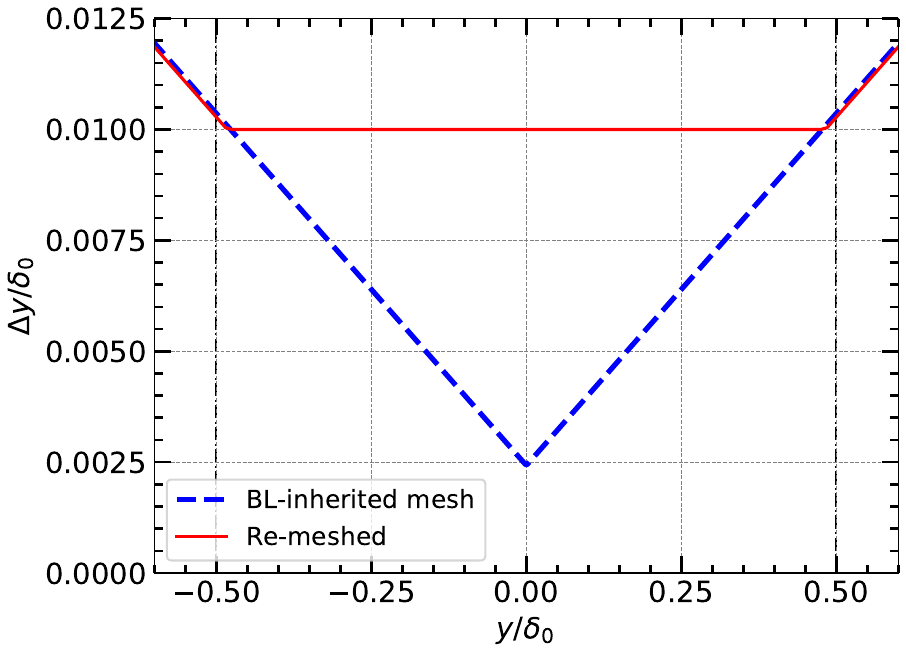}
  \caption{Vertical grid-spacing in the core of the shear layer, comparing the mesh inherited from the precursor boundary layers and used
  for $t^* \leq 0.75$ (blue dashed), and after re-meshing at $t^*=0.75$ (red solid).}
\label{slDeltaY}
\end{figure}

\label{secSLverticalgrid}
The vertical grid-spacing $\Delta y$ around $y=0$ is set by the boundary layer requirement of resolving the viscous sublayer; this spacing is unnecessarily fine for the shear layer simulations once the initial wake has closed.
We therefore run the shear layer on the mesh inherited from the boundary layer simulations only for $t^* \leq 0.75$ and then interpolate the solution onto a mesh with uniform vertical resolution in the middle of the shear layer; the grid-spacings of both meshes are shown in figure~\ref{slDeltaY}.
The choice of $t^*=0.75$ and the new smallest grid-spacing $\Delta y/\delta_\mathrm{0} = 0.01$ was aided by using the LES error estimation method of~\citet{Toosi_CF_2017}.

\begin{table}
  \begin{center}
  \caption{Description of the precursor boundary layers simulated using LES to set up the initial condition for the 
  shear layer simulations. Reynolds numbers are defined as $Re_\tau = u_\tau \deltaTBL/\nu$ and $Re_\theta = \|\boldvec{U}\|\theta_\text{TBL}/\nu$.
  Grid resolutions are reported in the boundary layer inner scaling.}
  \label{tableTemporalBLcases}
  \begin{tabular}{ccccccccl}
   Case &$\|\boldvec{U}\|/\|\boldvec{U}_\text{L}\|$ &$\theta$ & $Re_\tau$ & $Re_{\theta}$
   &$\Delta y^+_w$&$\Delta x^+, \Delta z^+$\\
  BLU100D00 ($\tbl{L}$)& $1$ & $0^\circ$ & 246 & 569 & 1.2 & 10.3  \\ 
  BLU150D00 ($\tbl{H}$-planar)& $1.5$ & $0^\circ$ & 347 & 846 & 1.7 & 14.4 \\
  BLU117D25 ($\tbl{H}$-skewed)& $1.17$ & $25^\circ$ & 288 & 682 & 1.4 & 12 \\
  \end{tabular}
\end{center}
\end{table}

\subsection{Solution verification}

The grid resolution for boundary layer simulations using the \texttt{Hybrid} code with $\Delta x=\Delta z$ and grid-spacings as listed in Table~\ref{tableTemporalBLcases} was assessed and found sufficient in~\citet{Larsson_AIAA_2022}.
In addition, the present boundary layer simulations are validated against the DNS results of \citet{Schlatter_JFM_2010} in Appendix~\ref{appendixBL}.

\begin{figure}
  \centering

  \begin{subfigure}[t]{0.47\columnwidth}
      \centering
      \includegraphics[width=\linewidth]{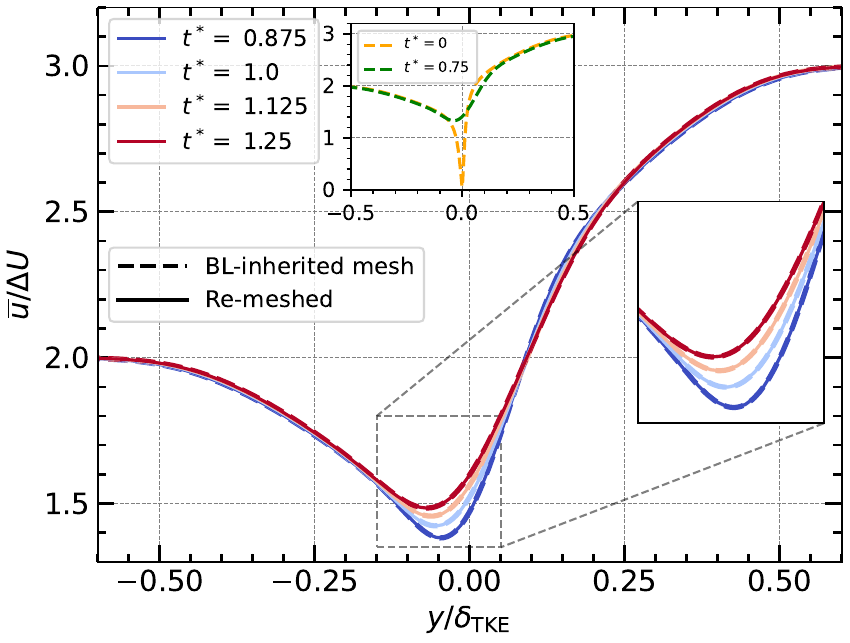}
  \end{subfigure}
  \hfill
  \begin{subfigure}[t]{0.49\columnwidth}
      \centering
      \includegraphics[width=\linewidth]{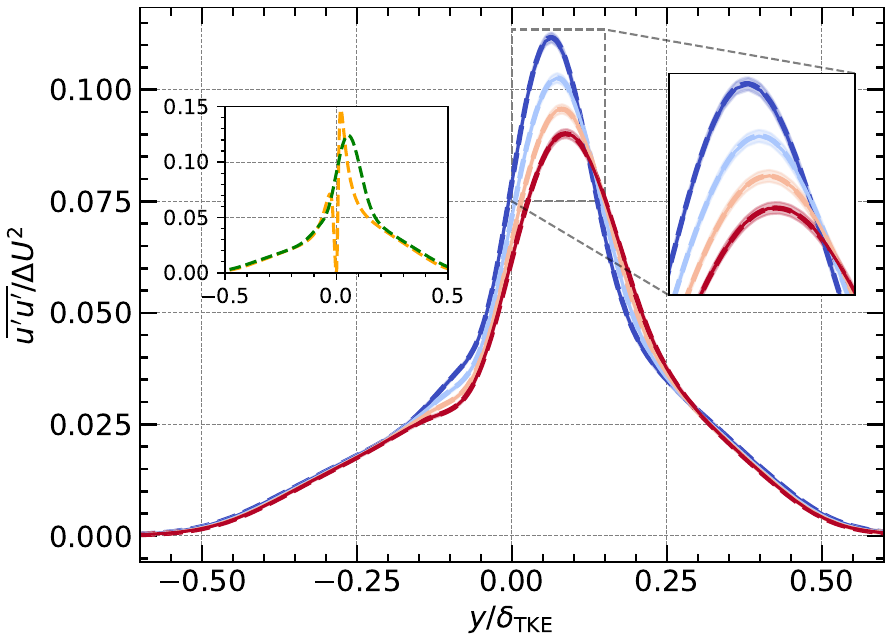}
  \end{subfigure} 

  \caption{Influence of the re-meshed vertical grid-spacing $\Delta y$ assessed by comparing results for the planar shear layer on the re-meshed grid versus a grid kept fine for all times.}
  \label{slGridConvProfiles}
\end{figure}

The shear layers are initialized from the boundary layers using the same grid; since the turbulence length scales grow in the shear layer, this means that the shear layer grids are adequately resolved initially and overly resolved at later times.
The only possible exception is the vertical grid-spacing $\Delta y$ since this is re-meshed at $t^* = 0.75$ as discussed in section~\ref{secSLverticalgrid}.
To test the effect of this choice, simulations are compared with and without the re-meshing (i.e., maintaining the boundary layer inherited fine $\Delta y$ mesh for later times) up to $t^* = 1.25$.
Some key results are shown in figure~\ref{slGridConvProfiles}, showing minimal effect of the re-meshing.

\begin{figure}
  \centering
  \includegraphics[width=0.6\textwidth]{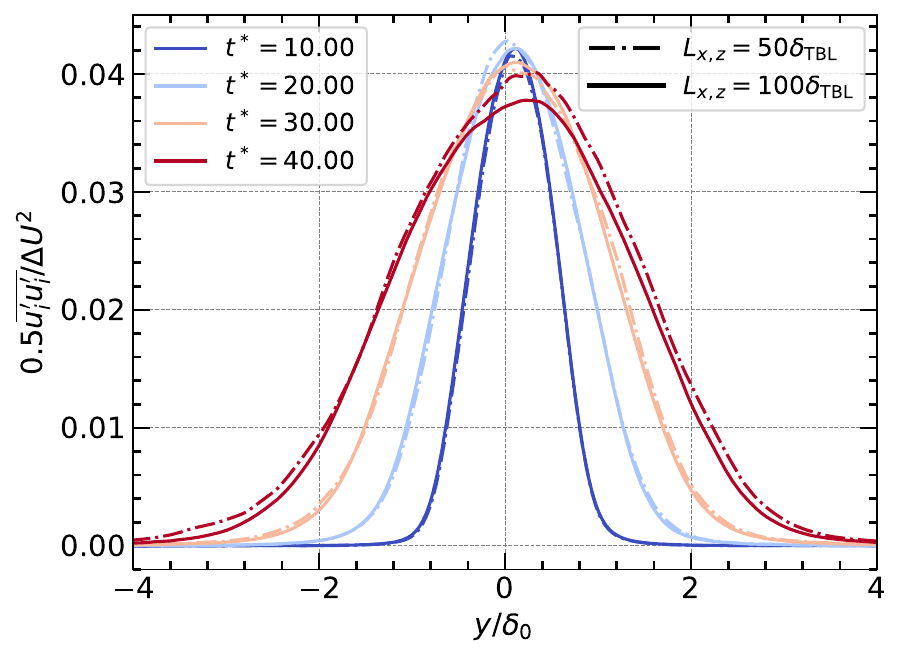}
  \caption{Turbulence kinetic energy profiles for the skewed shear layer on the baseline 
  domain ($L_{x,z} = 100 \deltaTBL$) and a domain of half the size
  ($L_{x,z} = 50 \deltaTBL$).}
\label{tkeDomainDependencePlot}
\end{figure}

\begin{figure}
  \centering
  \begin{subfigure}{0.49\columnwidth}
      \centering
      \includegraphics[width=\linewidth]{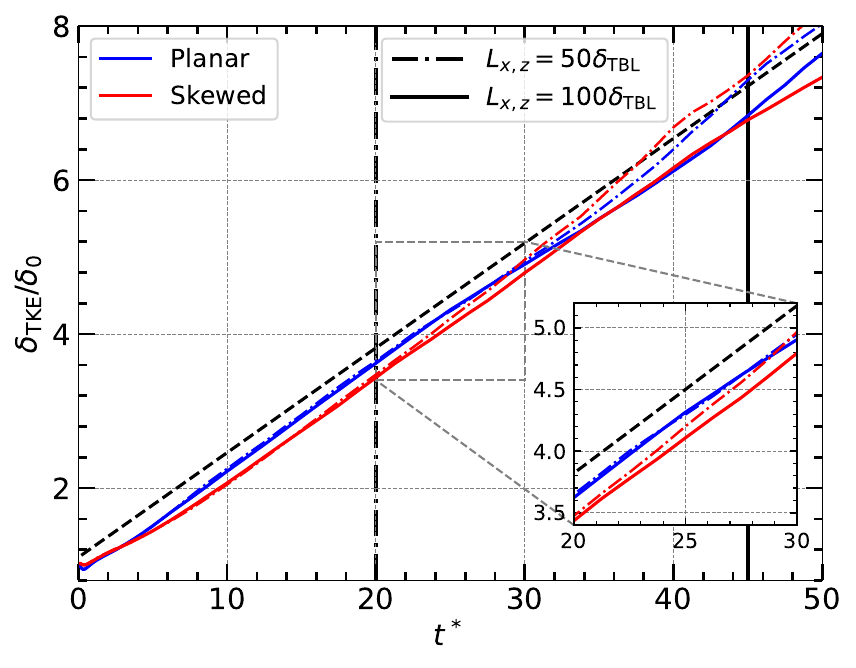}
  \end{subfigure}
  \hfill
  \begin{subfigure}{0.49\columnwidth}
      \centering
      \includegraphics[width=1.06\linewidth]{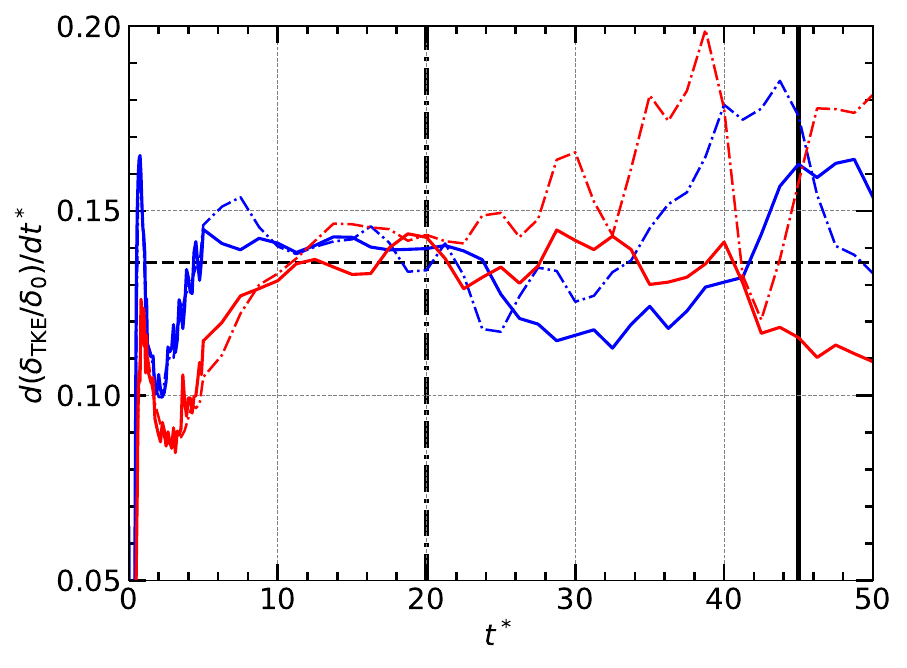}
  \end{subfigure} 
  \caption{Shear layer thickness $\deltaTKE$ (left) and its growth rate (right) for the planar (blue) and skewed (red)
  shear layers on the baseline domain ($L_{x,z} = 100 \deltaTBL$, solid lines) and a domain of half the size
 ($L_{x,z} = 50 \deltaTBL$, dash-dot lines).
 The black dashed lines represents linear growth ($\deltaTKE/\delta_\mathrm{0} = 0.136t^* + 1.1$ in the left subplot), whereas the black vertical lines mark the approximate 
  time $t^*$ when the shear layers exhibit finite domain size effects on their respective domain sizes.}
  \label{deltaTKEplot}
\end{figure}

The computational domain size will eventually influence the solution by limiting the evolution of the largest flow structures, resulting in a maximum $t^*$ for which the results are meaningful.
This time is estimated by running additional simulations in domains that are half as large in the horizontal directions (i.e., $L_{x,z}/\deltaTBL = 50$),
with sample results shown in figures~\ref{tkeDomainDependencePlot} and~\ref{deltaTKEplot}.
The results are very similar on the two grids for $t^* \lesssim 20-22$ at which point 
$\deltaTKE / \delta_\mathrm{0} \approx 3.5$.
We thus estimate that the results are independent of the domain size while 
$L_{x,z}/\deltaTKE 
= (L_x/\deltaTBL) / (2 \deltaTKE/\delta_\mathrm{0})
\lesssim 50 / (2\cdot 3.5) \approx 7$.
The twice larger baseline domain should then be accurate until $\deltaTKE/\delta_\mathrm{0} \approx 7$, which translates to $t^* \approx 45$.

\section{Impact of skew on mean flow and variances}
We now assess the impact of skew on the shear layer evolution, initially focusing on the mean flow and variances.
We start by noticing that both the planar and skewed shear layers exhibit essentially the same linear growth rate
$d(\deltaTKE/\delta_0)/dt^* = d\deltaTKE/d(t \Delta U) \approx 0.136$
(indicated by the black dashed line) 
for $t^* \gtrsim 10$ in figure~\ref{deltaTKEplot},
but that the skewed case grows a bit more slowly than the planar case for $t^* \lesssim 10$.


\subsection{Mean velocities in the fixed lab frame}\label{subSecVelLabFrame}
\begin{figure}
  \centering
  \begin{subfigure}[t]{0.49\columnwidth}
      \centering
      \includegraphics[width=\linewidth]{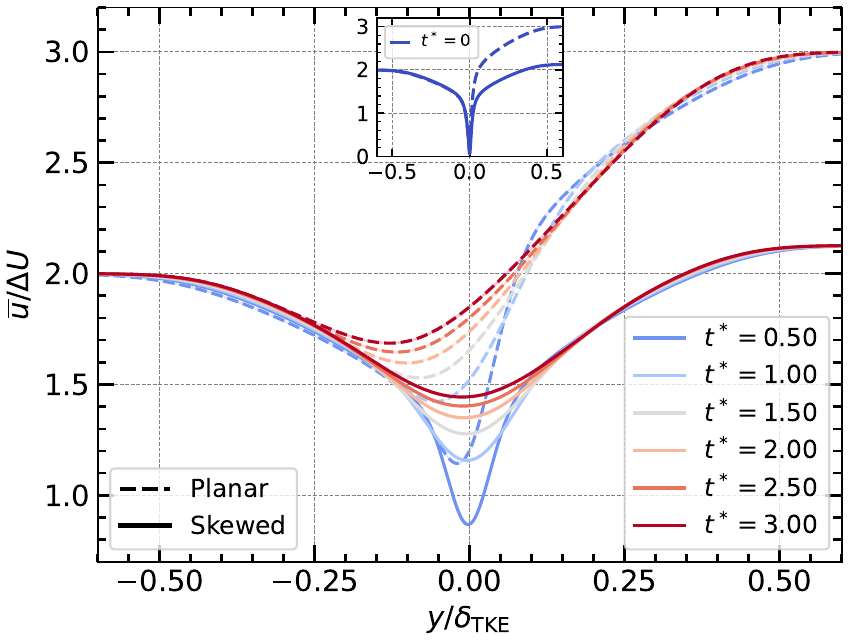}
  \end{subfigure}
  \hfill
  \begin{subfigure}[t]{0.49\columnwidth}
      \centering
      \includegraphics[width=\linewidth]{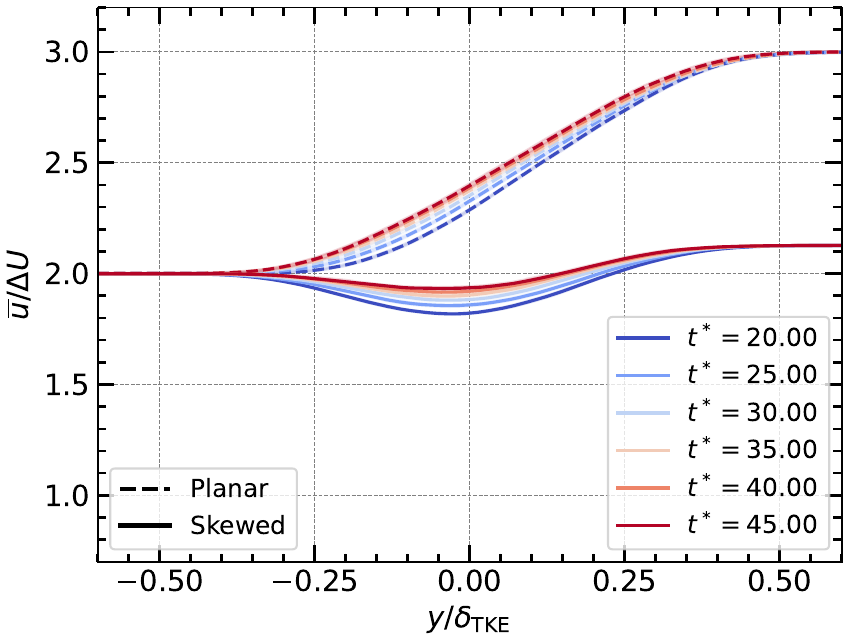}
  \end{subfigure} 
 
  \begin{subfigure}[t]{0.49\columnwidth}
      \centering
      \includegraphics[width=\linewidth]{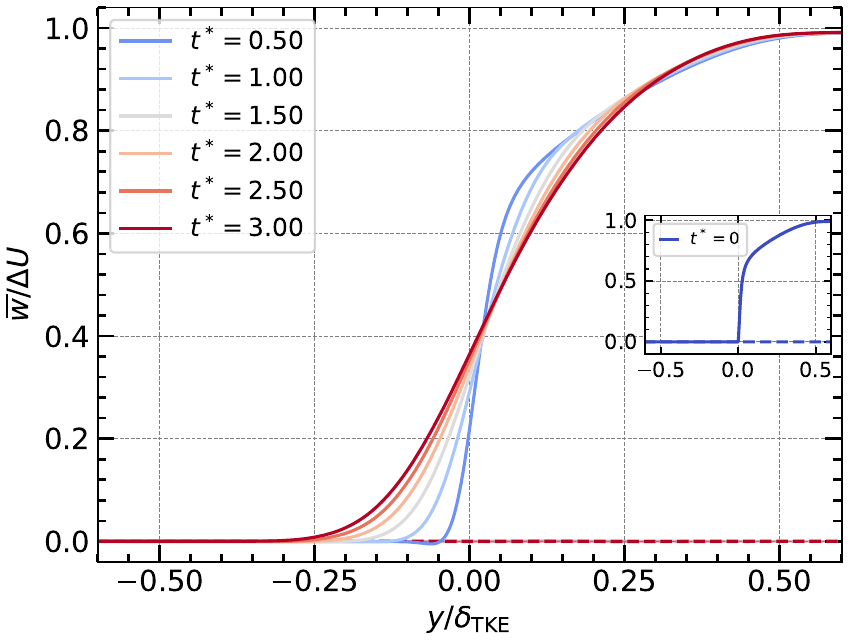}
      \caption{Initial development.}
  \end{subfigure}
  \hfill
  \begin{subfigure}[t]{0.49\columnwidth}
      \centering
      \includegraphics[width=\linewidth]{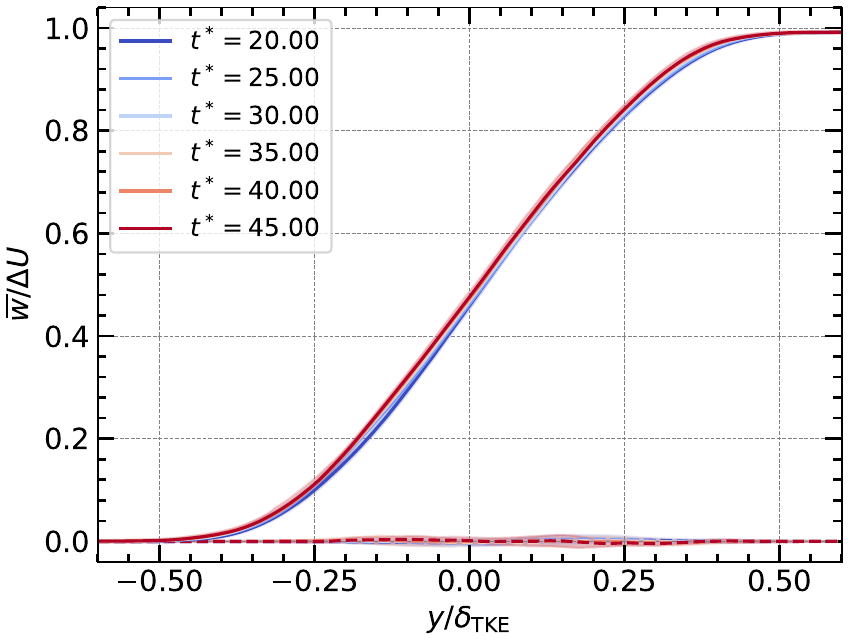}
      \caption{Later development.}
  \end{subfigure} 
  \caption{Mean velocity profiles in the lab frame of reference 
  for the planar (dashed lines) and skewed (solid lines) shear layers.
  The insets show the initial conditions at $t^* = 0$.
  Shaded areas represent the $95\%$ confidence interval due to averaging.}
  \label{velocityProfilesLabFrame}
\end{figure}

The planar and skewed cases are compared first in the lab frame of reference, meaning velocity components aligned with the imagined wind tunnel or flow direction; this is shown in 
figure~\ref{velocityProfilesLabFrame}.
Starting with $\widebar{u}$, the top-left figure shows that the large initial
momentum deficit due to the prescribed initial conditions (shown in the inset plot) decays rapidly during the 
early transient. The deficit decays faster for the planar case, possibly due to the sharper initial velocity
gradient at the interface. At later times (top-right figure), both cases approach a quasi-equilibrium state 
which differs on the high-speed side due to a mismatch in $\boldvec{U}_\text{H}$. 

The behavior of $\widebar{w}$ further highlights this difference, since it remains essentially zero for the 
planar case (except for minor variations due to finite averaging errors), while the skewed case settles into a finite 
non-zero profile. Together, these observations suggest that a comparison limited to the fixed lab frame may incorrectly imply that three-component effects are persistent in shear layers.

\subsection{Mean velocities in the convected mean shear frame}\label{subSecVelMNframe}
\begin{figure}
  \centering
  \begin{subfigure}{0.49\columnwidth}
      \centering
      \includegraphics[width=\linewidth]{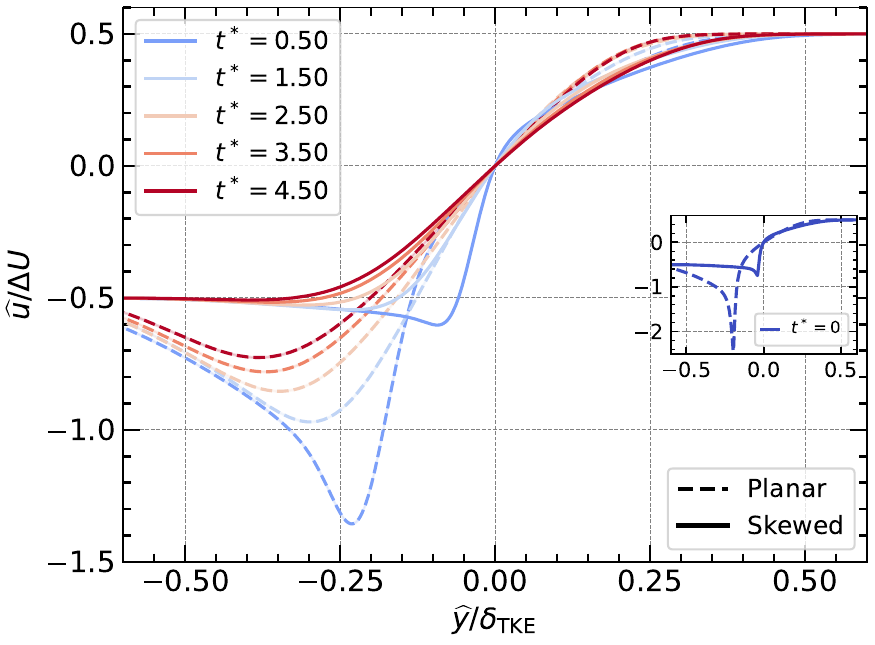}
      \caption{Initial development.}
  \end{subfigure}
  \hfill
  \begin{subfigure}{0.49\columnwidth}
      \centering
      \includegraphics[width=\linewidth]{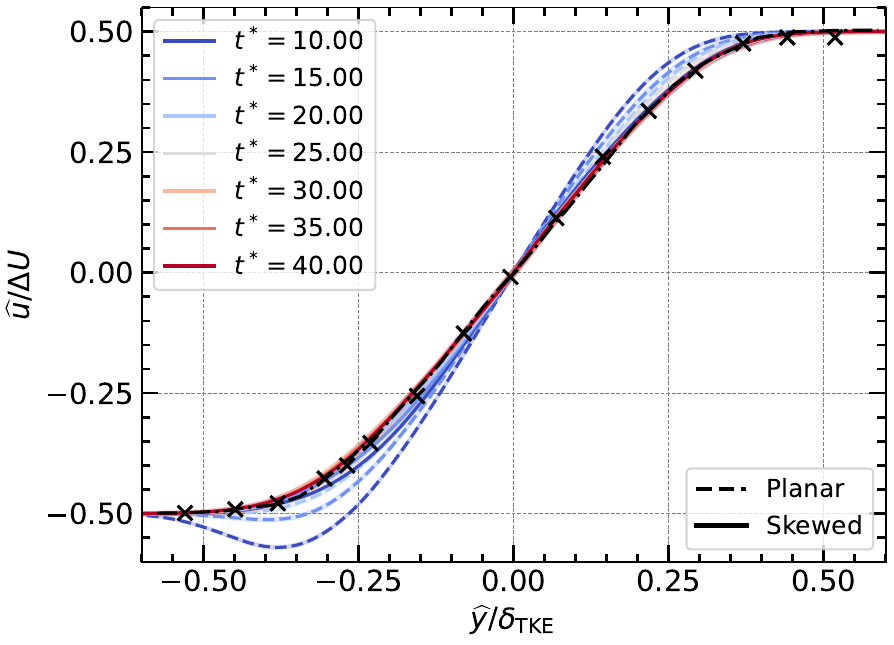}
      \caption{Later development.}
  \end{subfigure} 
  \caption{Mean velocity profiles in the direction of the mean shear, i.e.,
  $\widehat{u}$ along $\boldvec{m}$
  for the planar (dashed lines) and skewed (solid lines) shear layers.
  The inset shows 
  the initial condition at $t^*=0$.
  Shaded areas represent the $95\%$ confidence interval.
  Digitized data of \citet{Bell_AIAA_1990} (black `$\times$' symbols) and \citet{Rogers_PoF_1994} (black dash-dot line) 
  are overlaid for comparison in the right subplot.}
  \label{uHatProfiles}
\end{figure}

We start by comparing the evolution of the mean $\widehat{u}$ profiles (along $\boldvec{m}$) in figure~\ref{uHatProfiles}.
The results are divided into two figures to highlight two distinct flow regimes. The left figure shows an early transient stage where the 
initial momentum deficit in $\widehat{u}$ (shown in the inset plot) decays rapidly with time. Similar to $\widebar{u}$, the deficit decay rate
is case specific, and a collapse between planar and skewed cases is not expected at this stage. Nonetheless, unlike the fixed lab frame, the 
flow in both the planar and skewed shear layers is bounded by the same freestream values ($\widehat{u}/|\Delta U| = \pm 0.5$). Therefore, 
when comparing the evolution at later times in the right figure, the profiles for both planar and skewed shear layers collapse into identical self-similar 
states that are characteristic of a canonical shear layer. This is verified by the agreement between present results and digitized data of 
\citet{Bell_AIAA_1990} (black `$\times$' symbols) and \citet{Rogers_PoF_1994} (black dash-dot line) in the right subplot.
 
\begin{figure}
  \centering

  \begin{subfigure}[t]{0.49\columnwidth}
      \centering
      \includegraphics[width=\linewidth]{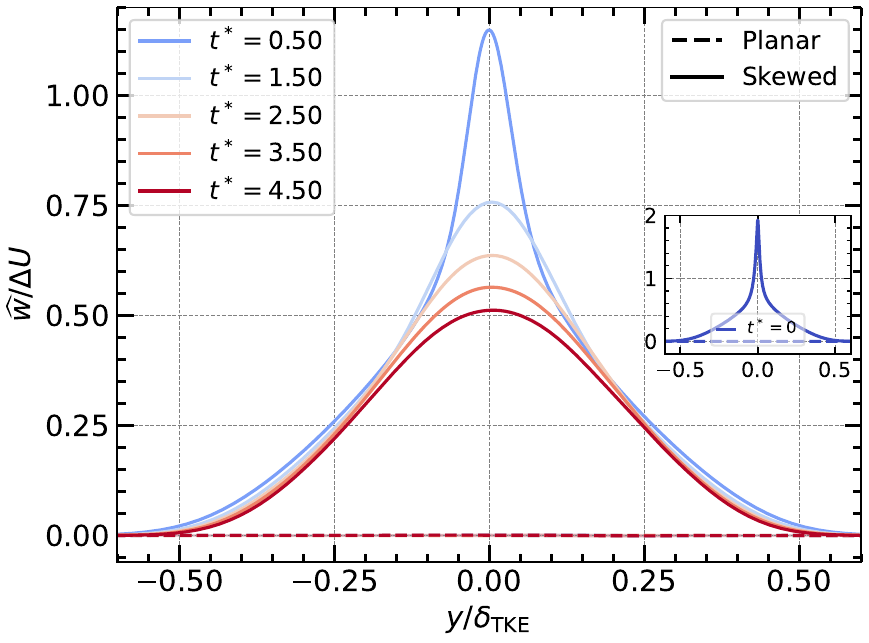}
      \caption{Initial development.}
  \end{subfigure}
  \hfill
  \begin{subfigure}[t]{0.49\columnwidth}
      \centering
      \includegraphics[width=\linewidth]{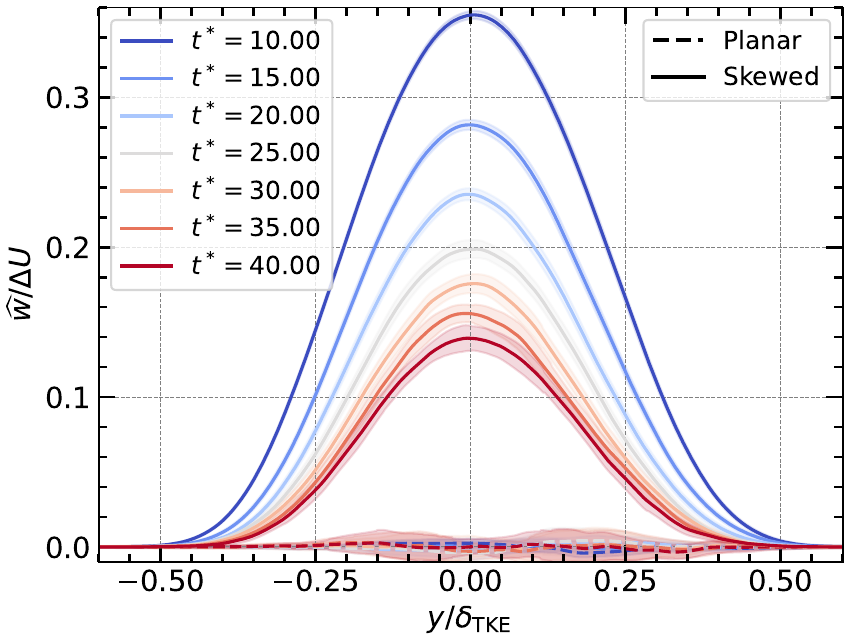}
      \caption{Later development.}
  \end{subfigure} 
  \caption{Mean velocity profiles in the direction normal to the mean shear, i.e, 
  $\widehat{w}$ along $\boldvec{n}$.
  The inset shows the initial condition at $t^*=0$.
  Shaded areas represent the $95\%$ confidence intervals.}
  \label{wHatProfiles}
\end{figure}

We next look at the evolution of $\widehat{w}$ 
(along $\boldvec{n}$) 
in figure~\ref{wHatProfiles}.
Note that $\widehat{w}$ should be zero for the planar shear layer,
with the minor deviations being due to finite averaging. 
For the skewed shear layer, $\widehat{w}$
does not show a qualitative change over time like $\widehat{u}$. Instead, the jet-like profiles
continue to decay monotonically throughout the entire simulation timeframe. Importantly, 
$\widehat{w}$ is finite even when $\widehat{u}$ for the skewed shear layer appears to evolve self-similarly 
like the planar shear layer. In addition, we make two observations. First, $\widehat{w}$ profiles are symmetric about the splitter plate 
plane $y = 0$ instead of $\widehat{y}=0$, the long-time symmetry plane of $\widehat{u}$. Second, $\widehat{w}$ profiles plotted against the scaled 
coordinate $y/\deltaTKE$ do not collapse along their width. 

\begin{figure}
  \centering
  \begin{subfigure}[t]{0.49\columnwidth}
      \centering
      \includegraphics[width=1\linewidth]{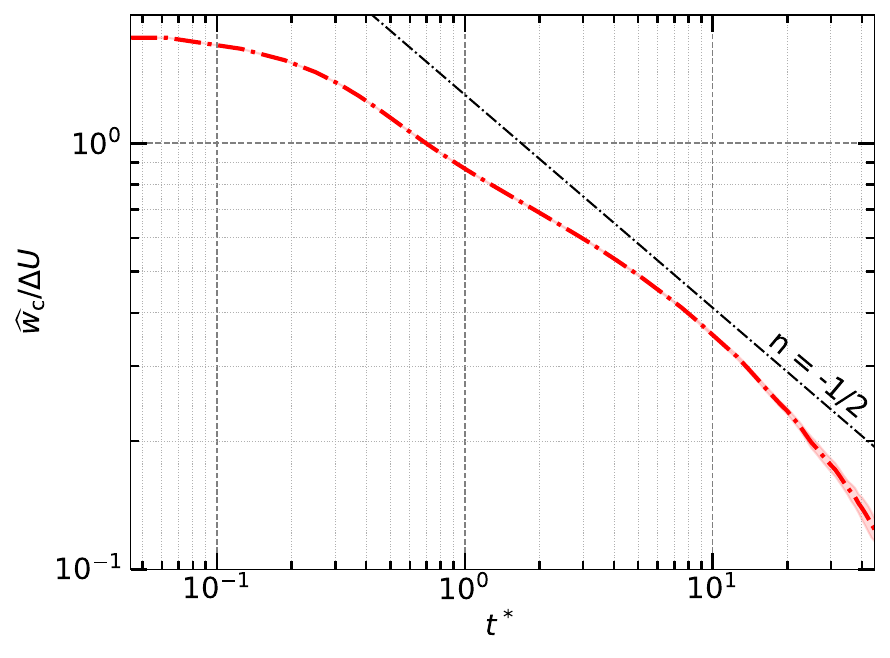}
  \end{subfigure}
  \hfill
  \begin{subfigure}[t]{0.49\columnwidth}
      \centering
      \includegraphics[width=0.9\linewidth]{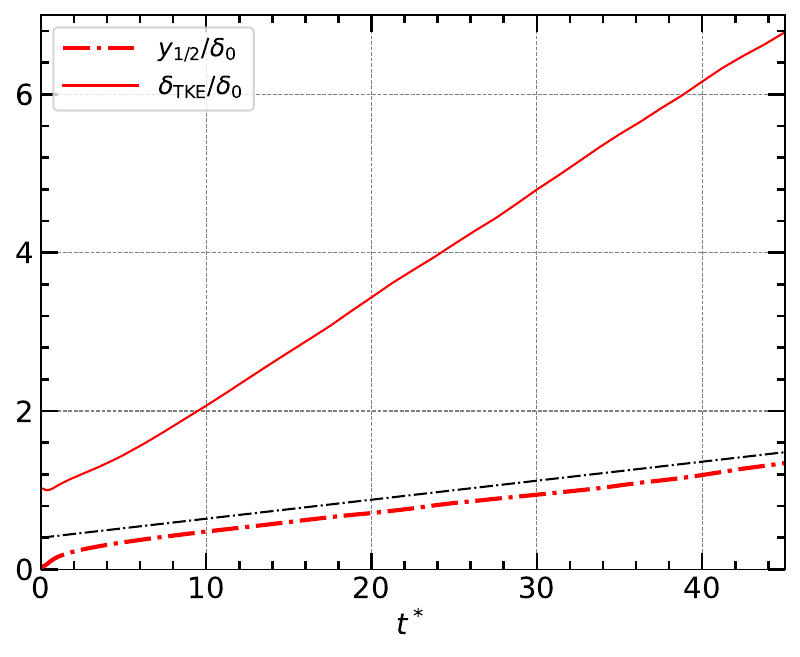}
  \end{subfigure} 
  \caption{Evolution of the jet centerline velocity $\widehat{w}_c/\Delta U$ (left) and 
  jet half-width $y_{1/2}/\delta_\mathrm{0}$ (right) by treating $\widehat{w}$ of the skewed shear layer as
  a planar jet. The black dash-dotted lines represent
  power law decay with exponent $n=-1/2$ and linear growth ($y_{1/2}/\delta_\mathrm{0} = 0.024t^* + 0.4$), respectively.
  The shaded area in the 
  left subplot represents the $95\%$ confidence interval.}
  \label{jetProperties}
\end{figure}

These observations suggest that the evolution of $\widehat{w}$ does not follow the 
self-similar scaling for the shear layers. Instead, in the absence of body forces or an imposed shear along $\boldvec{n}$, the evolution of $\widehat{w}$ is primarily 
driven by turbulent and viscous diffusion, leading to a jet-like evolution. To assess this, we treat $\widehat{w}$ as a planar jet and 
examine the evolution of the jet centerline velocity $\widehat{w}_c(t^*) = \widehat{w}(y=0,t^*)$, and jet half-width $y_{1/2}$ defined as 
$\widehat{w}(y=y_{1/2},t^*) = 0.5 \widehat{w}_c(t^*)$. The results are shown in 
figure~\ref{jetProperties} (red dash-dot lines). For $t^* \gtrsim 5$, they approximately agree with the expected self-similar scaling of a planar jet 
($\widehat{w}_c \sim t^{*-1/2}$ and $y_{1/2} \sim t^*$, marked as black dash-dot lines) derived from self-similarity arguments \citep{Pope_2000}.
The growth rate of this planar jet is $d(y_{1/2}/\delta_\mathrm{0})/dt^* \approx 0.024$, which is significantly lower (by a factor of $\sim 5$) than that of $\deltaTKE$ which is also shown in the figure for visual comparison.

\begin{figure}
  \centering
  \begin{subfigure}[t]{0.49\columnwidth}
      \centering
      \includegraphics[width=\linewidth]{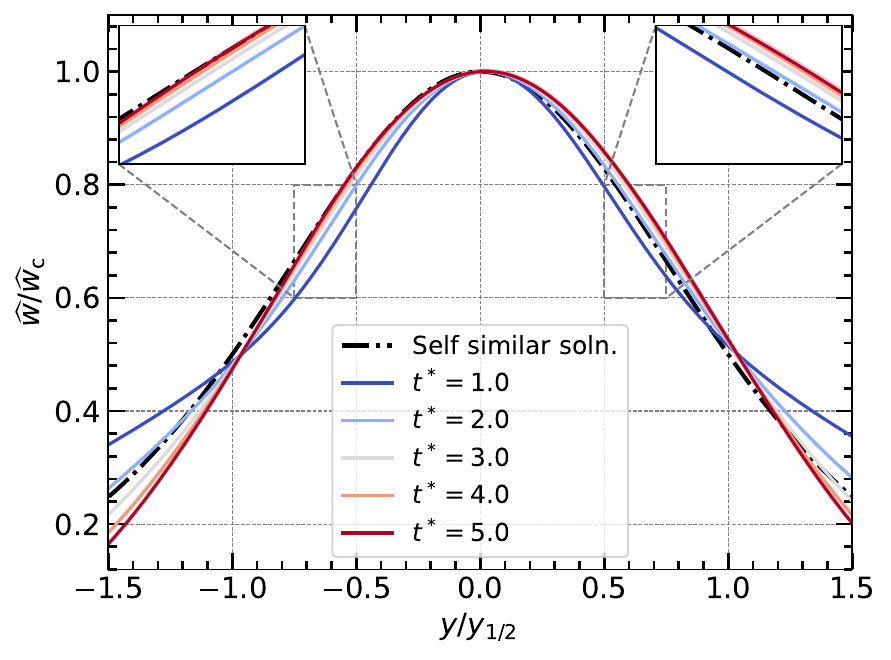}
      \caption{Initial development.}
  \end{subfigure}
  \hfill
  \begin{subfigure}[t]{0.49\columnwidth}
      \centering
      \includegraphics[width=\linewidth]{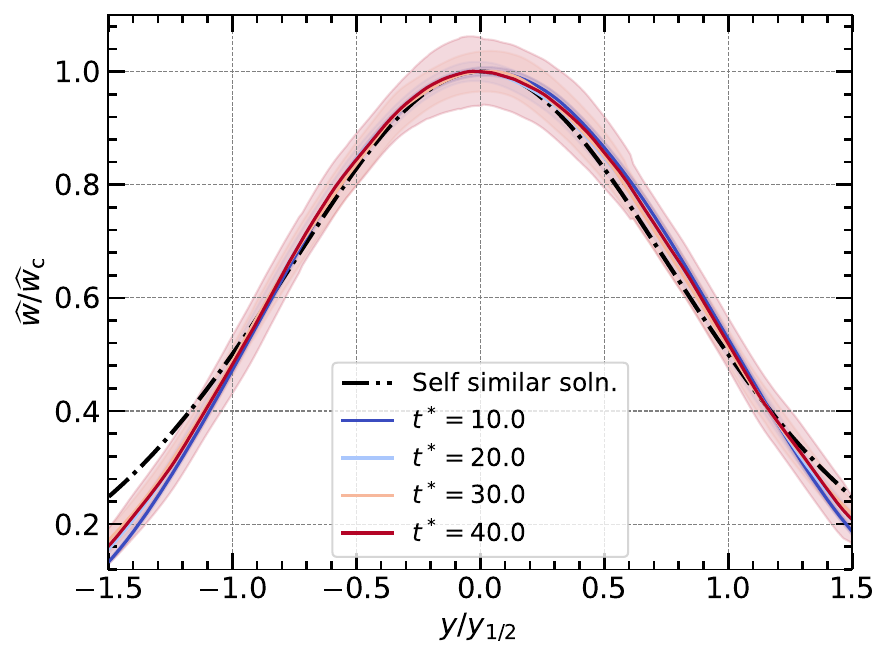}
      \caption{Later development.}
  \end{subfigure} 
  \caption{Mean velocity profiles in the direction normal to the mean shear, i.e, 
  $\widehat{w}$ along $\boldvec{n}$, plotted in planar jet scaling.
  The black dash-dot profile represents the self-similar solution for a planar jet.
  Shaded areas represent the $95\%$ confidence intervals.}
  \label{wHatJetProfiles}
\end{figure}

Using these quantities, we plot $\widehat{w}$ profiles in planar jet scaling 
as $\widehat{w}/\widehat{w}_c(y/y_{1/2})$ in figure~\ref{wHatJetProfiles}. For reference, we also overlay the expected self-similar solution for a planar 
jet \citep{Pope_2000}
\begin{equation}
  \widehat{w}/\widehat{w}_c = \sech^2(\alpha\; y/y_{1/2}),
  \  \alpha=0.5\ln \left(1+\sqrt{2} \right)^2
  ,
  \label{jetSelfSimilarEq} 
\end{equation}
as the black dash-dot lines. Similar to the trend observed in figure~\ref{jetProperties}, the scaled 
$\widehat{w}$ profiles overlap with the analytical definition for $t^* \gtrsim 5$, albeit with minor deviations for $y/y_{1/2} > 0$.
Overall, these results suggest that the temporal decay of $\widehat{w}$ indeed scales similarly to a planar jet, and that it is only weakly correlated
with the shear layer evolution. We note that this observation is consistent with the decoupling of the azimuthal (spanwise) component in
swirling jets \citep{Shiri_AIAA_2008}, which are analogous to skewed shear layers.

\subsection{Flow regimes}\label{subSecFlowRegimes}
\begin{figure}
  \centering
  \includegraphics[width=\columnwidth]{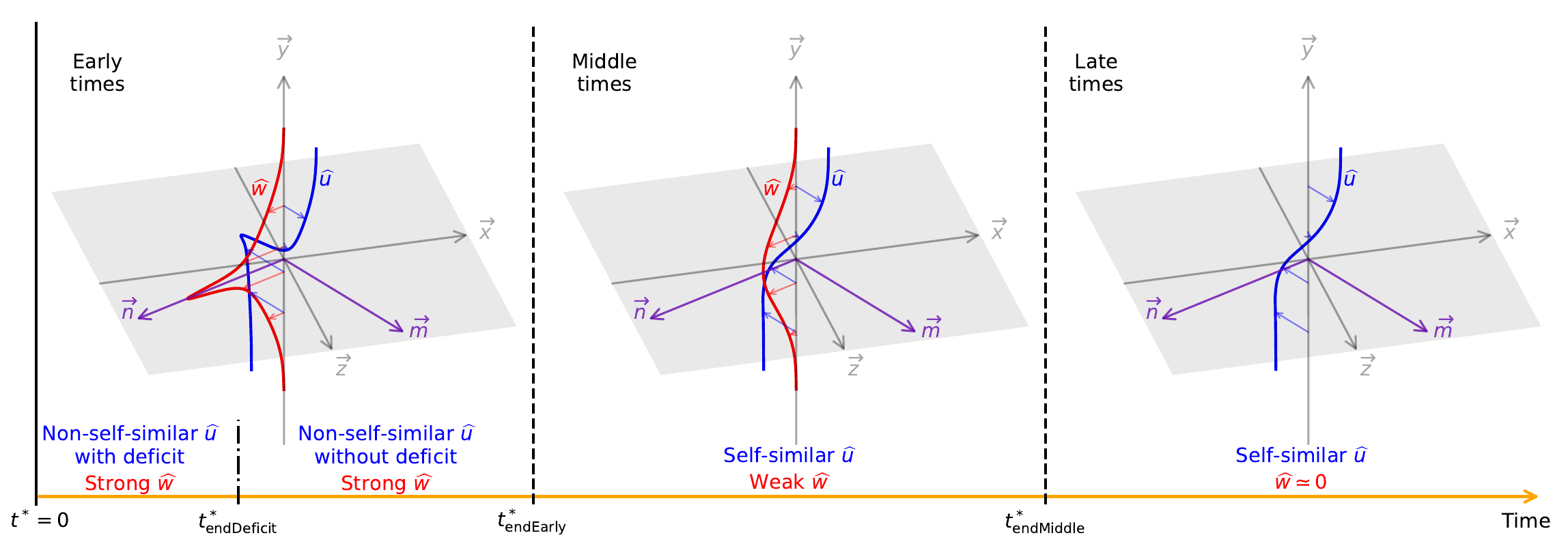}
  \caption{Flow regimes in skewed shear layers when viewed in a reference frame aligned with the mean shear direction (direction $\boldvec{m}$).
    }
\label{flowRegimesSchematic}
\end{figure}

Based on the qualitative nature of $\widehat{u}$ and $\widehat{w}$ profiles, the skewed shear layer evolution 
can be divided into three distinct flow regimes, illustrated in figure~\ref{flowRegimesSchematic}.
There is an ``early times'' regime during $0 < t^* < t^*_\text{endEarly}$, which is characterized by non-self-similar
evolution of $\widehat{u}$ and a decaying jet-like $\widehat{w}$.
This regime can be further subdivided at 
$t^*_\text{endDeficit}$ based on whether $\widehat{u}$ is monotonic or not (with a deficit from the initial condition).
Next, there is a ``middle times'' regime spanning $t^*_\text{endEarly} \leq t^* < t^*_\text{endMiddle}$
where $\widehat{u}$ evolves approximately self-similarly like a planar shear layer while $\widehat{w}$ decays approximately like a planar jet.
Finally, there is a ``late times'' regime for $t^* \geq t^*_\text{endMiddle}$ where $\widehat{w} \simeq 0$ and the skewed shear layer 
has become a planar one in the rotated reference frame.
Therefore, in the context of our hypothesis described in \S~\ref{subSecRefFrames},
three-component effects in a skewed shear layer should be limited to the early and middle times regimes.
For a planar shear layer, an early times regime can still be defined based on the evolution of $\widehat{u}$. However, with $\widehat{w} = 0$,
the flow transitions directly to the self-similar late times regime. 

\begin{figure}
  \centering
  \includegraphics[width=0.6\columnwidth]{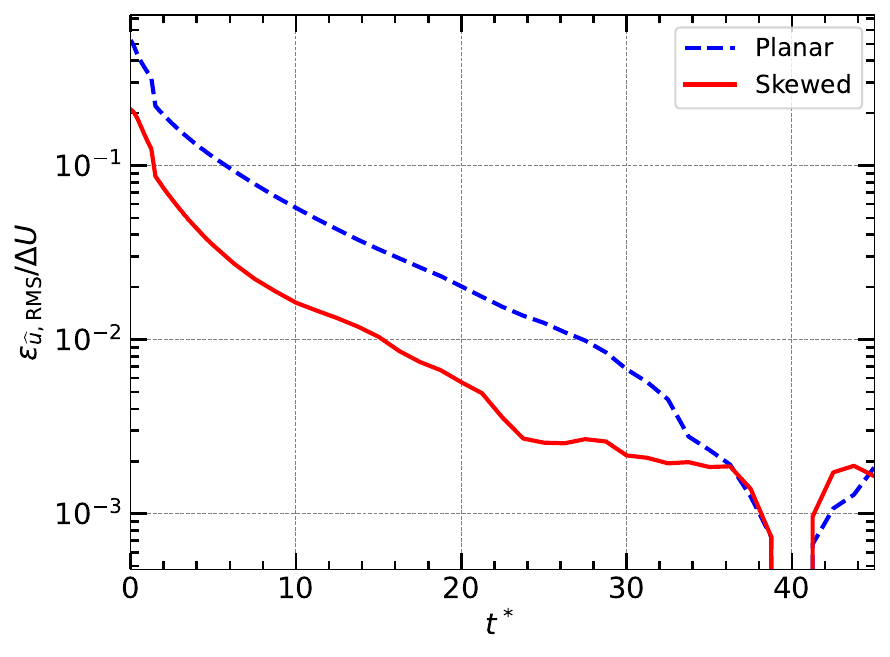}
  \caption{Root-mean-square (RMS) error in $\widehat{u}$ compared to its developed state at $t^* = 40$ 
  for the planar (blue dashed line) and skewed (red solid line) shear layers.}
\label{rmsErrorUhat}
\end{figure}

We can quantify the deviation in $\widehat{u}$ from its long-time self-similar profile as
\begin{equation}\label{rmsErrorUhatEq}
    \varepsilon_{\widehat{u},\mathrm{RMS}}(t^*) = \sqrt{\mathrm{mean}({\widehat{u}(t^*,\widehat{y}/\deltaTKE) - \widehat{u}_\mathrm{ref}(\widehat{y}/\deltaTKE)})^2},
\end{equation}
where $\widehat{u}_\mathrm{ref}$ is taken at $t^*=40$ from the respective cases, and the average is computed over $\widehat{y}/\deltaTKE \in [-1,1]$.
Figure~\ref{rmsErrorUhat} shows an overall monotonic decay of this deviation for both cases, indicating convergence towards the reference self-similar state.
Figure~\ref{uHatProfiles} visually suggests that $t^*_\text{endEarly} \approx 5-10$ for the skewed case, at which point the deviation from the reference state is about 2-3\% in figure~\ref{rmsErrorUhat}.
For the planar case, 
the same qualitative inspection of figure~\ref{uHatProfiles} 
and the same 2-3\% threshold in figure~\ref{rmsErrorUhat}
both suggest $t^*_\text{endEarly} \approx 20$.
The planar case does not have a ``middle times'' regime and goes straight to the late times, of course.

\begin{figure}
  \centering
  \includegraphics[width=0.6\columnwidth]{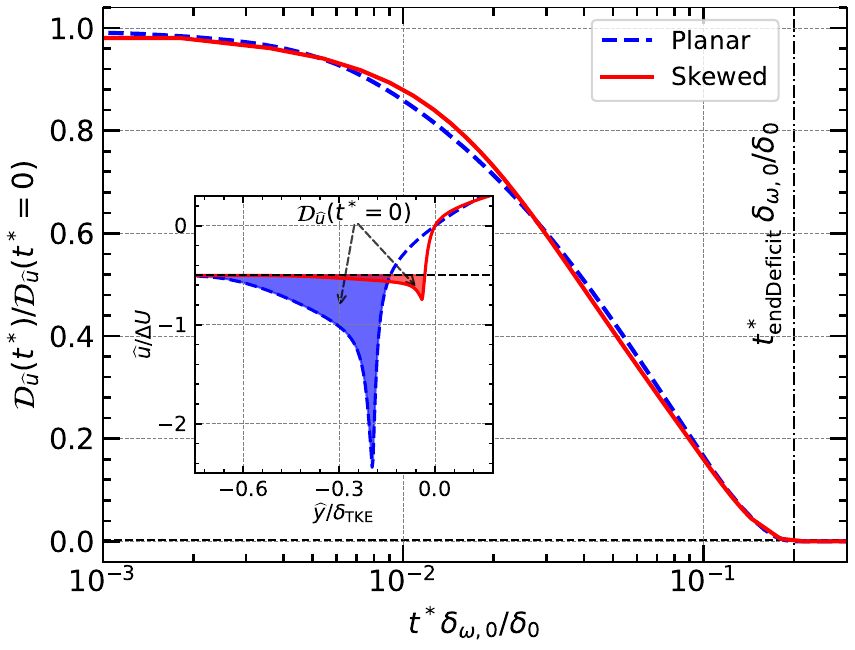}
  \caption{Decay of the integral momentum deficit $\mathcal{D}_{\widehat{u}}$ for the planar (blue dashed line)
  and skewed (red solid line) shear layers, scaled by the initial deficit $\mathcal{D}_{\widehat{u}}(t^*=0)$, and plotted against time 
  $t^*\delta_{\omega,o}/\delta_\mathrm{0}$, where $\delta_{\omega,o}$ is the initial shear layer vorticity thickness. The inset shows the 
  initial condition for $\widehat{u}$ for the two cases, with the shaded areas representing $\mathcal{D}_{\widehat{u}}(t^*=0)$.
  The black dash-dot line marks time $t^*_\mathrm{endDeficit}\delta_{\omega,o}/\delta_\mathrm{0} = 0.2$, where the deficit becomes negligibly small
  for both cases.}
\label{intDeficit}
\end{figure}

Within the early times regime, we also quantify $t^*_\text{endDeficit}$ by computing the integrated momentum deficit in $\widehat{u}$ as
\begin{equation}
\mathcal{D}_{\widehat{u}}(t^*) =
\int 
\max\!\left(0, -\left(0.5 + \dfrac{\widehat{u}(t^*,\widehat{y})}{\Delta U}\right)\right)
\, d \, \widehat{y}
\end{equation}
over $\widehat{y}/\deltaTKE \in [-1,1]$. The temporal decay of the normalized deficit $\mathcal{D}_{\widehat{u}}(t^*)/\mathcal{D}_{\widehat{u}}(t^*=0)$
is shown in figure~\ref{intDeficit}.
The initial deficit $\mathcal{D}_{\widehat{u}}(t^*=0)$ is represented by the shaded regions in the inset plot.
Note that the decay is plotted against the scaled time $t^*\delta_{\omega,\mathrm{0}}/\delta_\mathrm{0}$,
where $\delta_{\omega,\mathrm{0}} = \Delta U/|d\widehat{u}/dy|_{\mathrm{max},t^*=0}$ is the initial shear layer vorticity thickness.
With this scaling of the time, the decay of the momentum deficit behaves very similarly between the planar and skewed cases,
with the deficit vanished after only about 0.2 time units.
This translates to $t^*_\text{endDeficit} \approx 19$ and $7$ for the planar and skewed shear layers, respectively.
The observed collapse suggests
that the momentum deficit decay rate is primarily determined by the initial velocity gradient at the interface.

Next, for quantifying $t^*_\text{endMiddle}$, we note that $\widehat{w}$ for the skewed shear layer remains finite throughout the entire simulation 
(see figure~\ref{wHatProfiles}). Therefore, the flow does not reach the end of the middle times regime within the observable time 
window of the present study, leaving $t^*_\text{endMiddle}$ undetermined. Nonetheless, these results highlight a clear separation of 
timescales in a skewed shear layer.
The early times evolution towards an approximate self-similar state occurs over a relatively shorter timescale,
and it is governed by the prescribed initial conditions.
In contrast, the middle times evolution proceeds over a significantly longer timescale,
leading to gradually decaying three-component effects far downstream.

\subsection{Reynolds stress tensor in the convected mean shear frame}
\begin{figure}
  \centering

  \begin{subfigure}[t]{0.49\columnwidth}
      \centering
      \includegraphics[width=\linewidth]{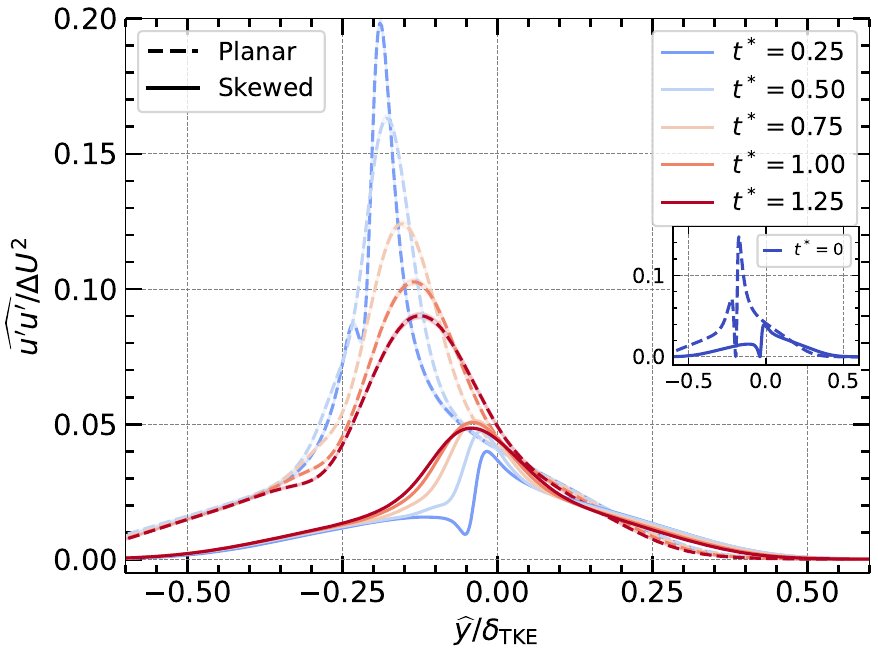}
  \end{subfigure}
  \hfill
  \begin{subfigure}[t]{0.49\columnwidth}
      \centering
      \includegraphics[width=\linewidth]{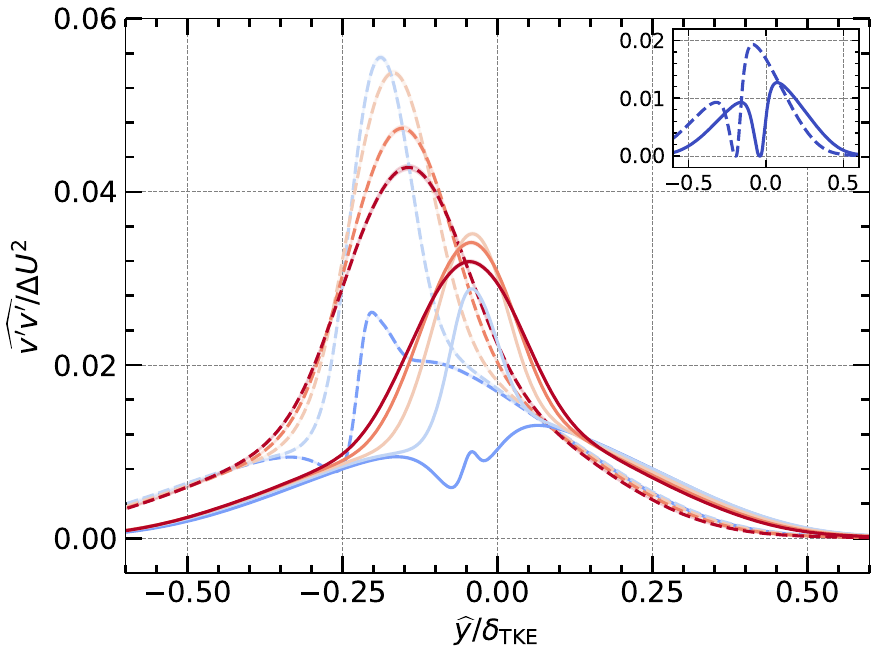}
  \end{subfigure} 

  \vspace{0.1truein}
 
  \begin{subfigure}[t]{0.49\columnwidth}
      \centering
      \includegraphics[width=\linewidth]{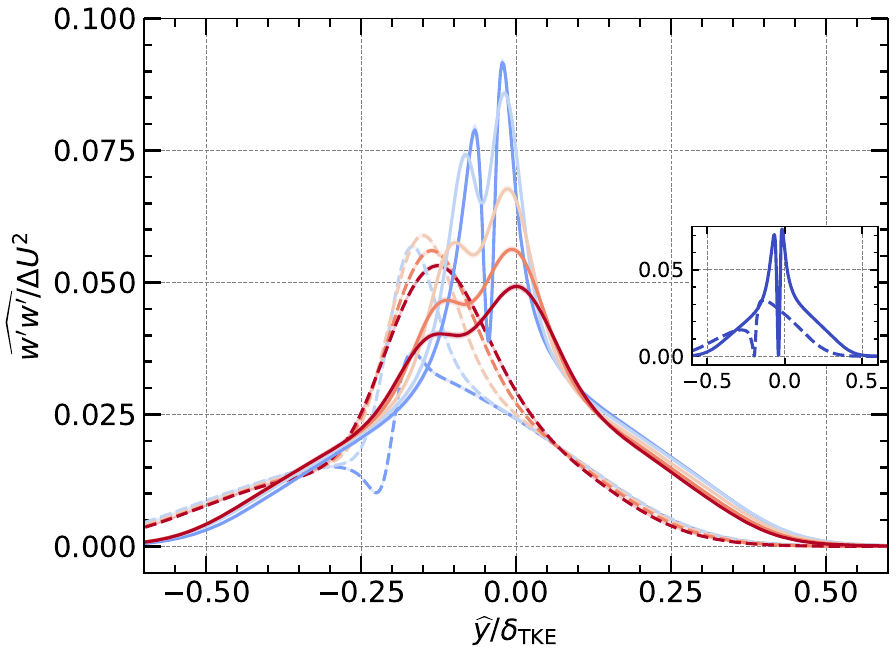}
  \end{subfigure}
  \hfill
  \begin{subfigure}[t]{0.49\columnwidth}
      \centering
      \includegraphics[width=\linewidth]{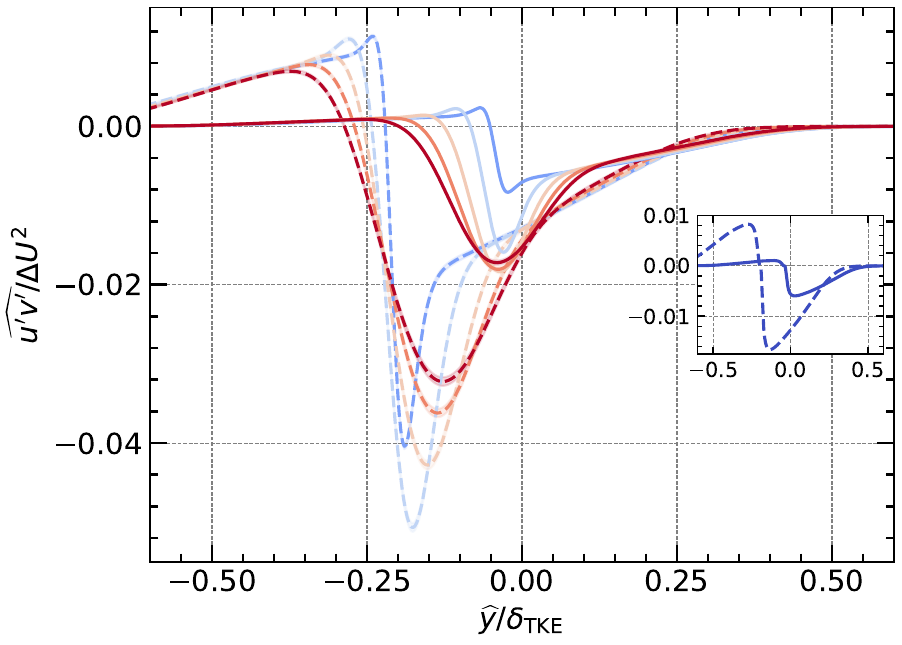}
  \end{subfigure}
 
  \vspace{0.1truein}
 
  \begin{subfigure}[t]{0.49\columnwidth}
      \centering
      \includegraphics[width=\linewidth]{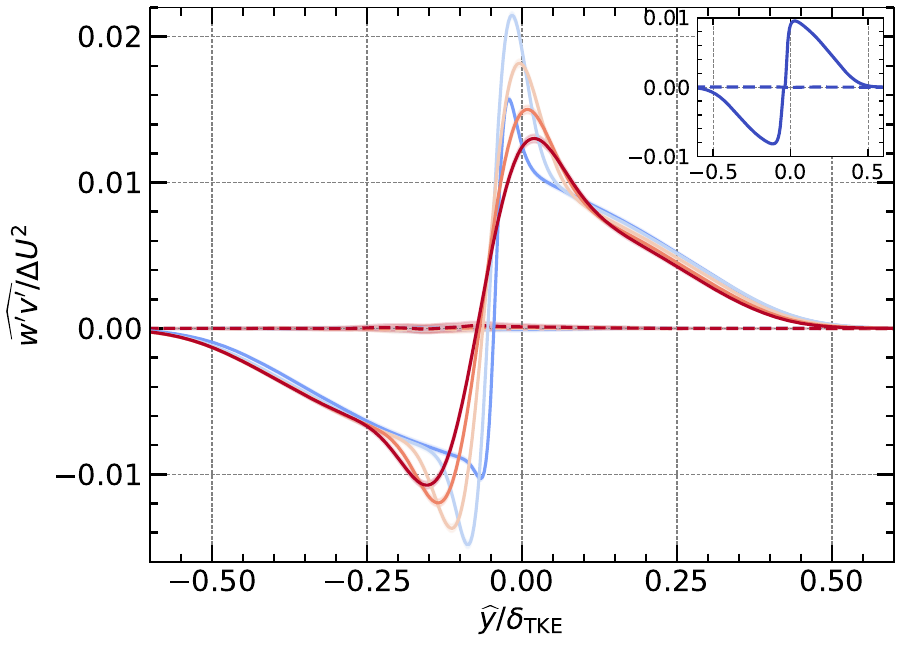}
  \end{subfigure}
  \hfill
  \begin{subfigure}[t]{0.49\columnwidth}
      \centering
      \includegraphics[width=\linewidth]{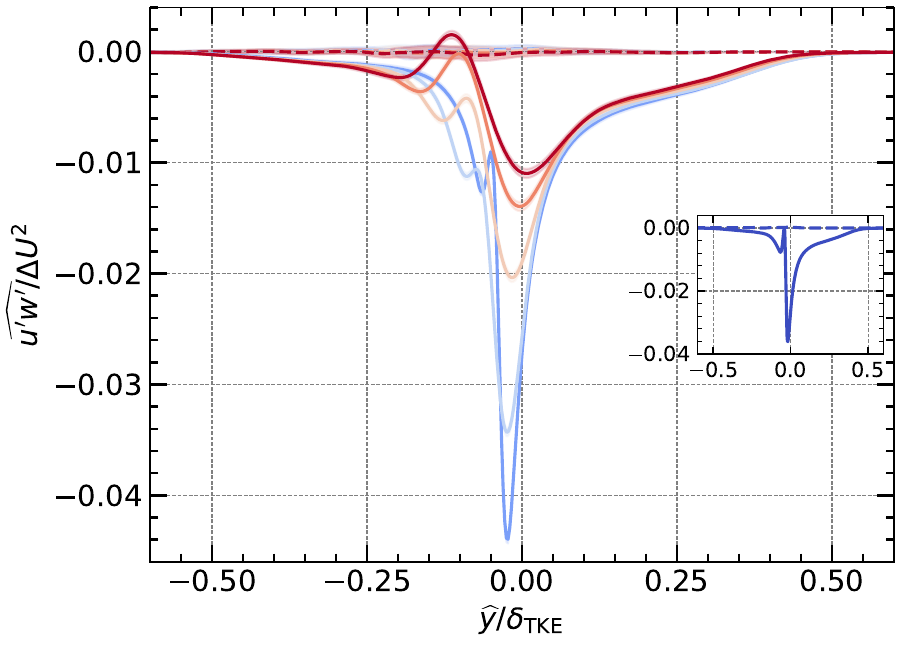}
  \end{subfigure}
  \caption{Reynolds stress tensor components in the rotated
  $(\boldvec{m},\boldvec{y},\boldvec{n})$ frame for the planar (dashed lines) and skewed (solid lines) shear layers
  during the early times regime. The inset shows the initial conditions for the respective components at $t^*=0$.
  The shaded areas represent the $95\%$ confidence intervals.}
  \label{ReStressProfilesInit}
\end{figure}

The Reynolds stress tensor rotated into the mean shear reference frame is denoted by $\widehat{u_i^\prime u_j^\prime}$.
All 6 components at the early times are shown in figure~\ref{ReStressProfilesInit}.
All components exhibit rapid growth from their boundary layer states during $0 < t^* \lesssim 0.5$, followed by gradual decay towards the middle times near-equilibrium-like state.
The remaining shear stress components $\widehat{v^\prime w^\prime}$ and $\widehat{u^\prime w^\prime}$ are nominally zero
for the planar case, and comparable to $\widehat{u^\prime v^\prime}$ for the skewed case in the early transient.

These observations are explained by the Reynolds stress production associated with the mean velocity gradients at the 
interface. The initially sharp gradients lead to strong production, while their subsequent diffusion reduces production over time
(see figures~\ref{uHatProfiles} and~\ref{wHatProfiles}). In particular, for the skewed shear layer, the weaker velocity gradient near 
the $\widehat{w}$ jet centerline leads to reduced Reynolds stress production, resulting in the relatively slower change in the shape of 
the $\widehat{w^\prime w^\prime}$ profile.

\begin{figure}
  \centering

  \begin{subfigure}[t]{0.49\columnwidth}
      \centering
      \includegraphics[width=\linewidth]{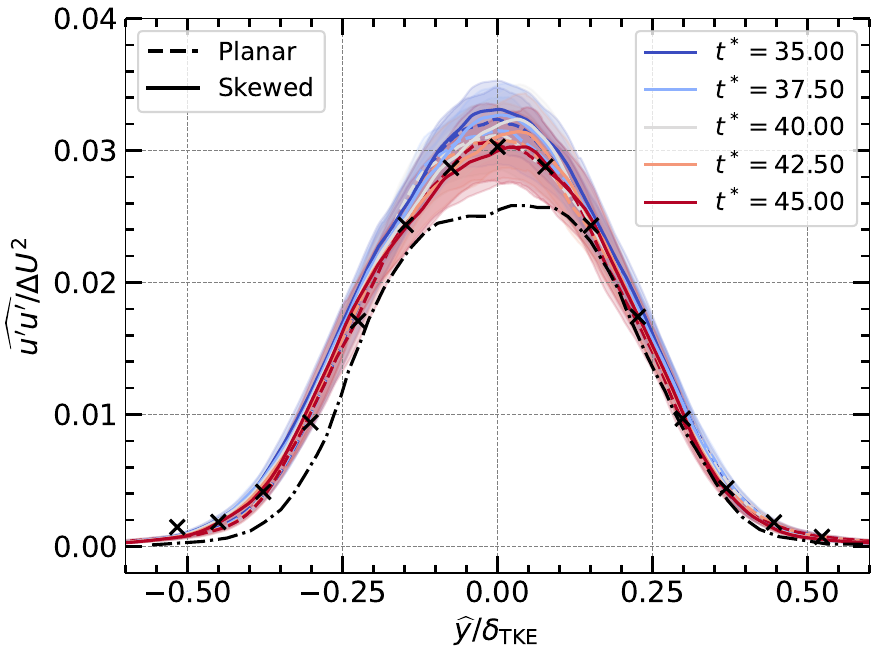}
  \end{subfigure}
  \hfill
  \begin{subfigure}[t]{0.49\columnwidth}
      \centering
      \includegraphics[width=\linewidth]{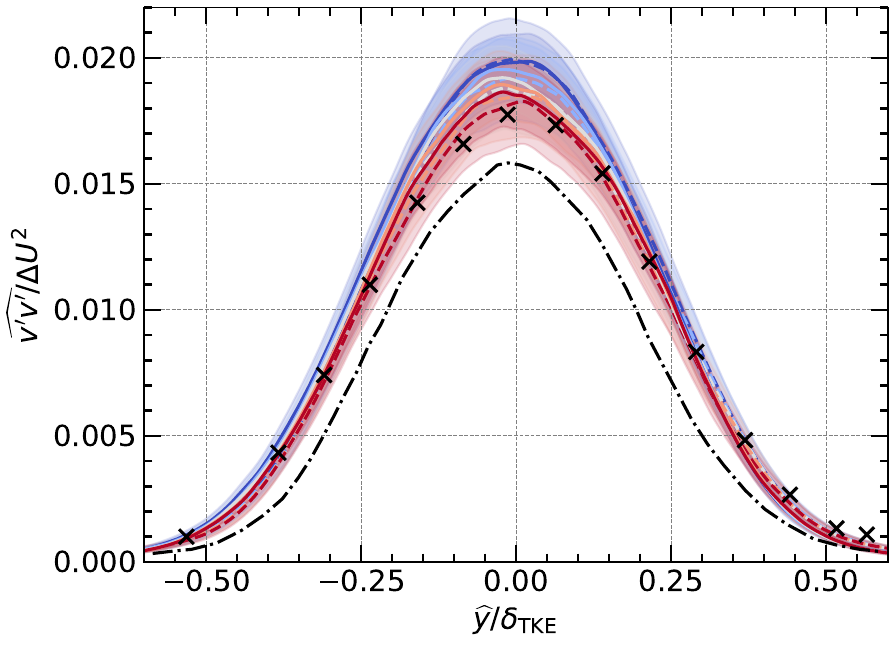}
  \end{subfigure} 

 
  \begin{subfigure}[t]{0.49\columnwidth}
      \centering
      \includegraphics[width=\linewidth]{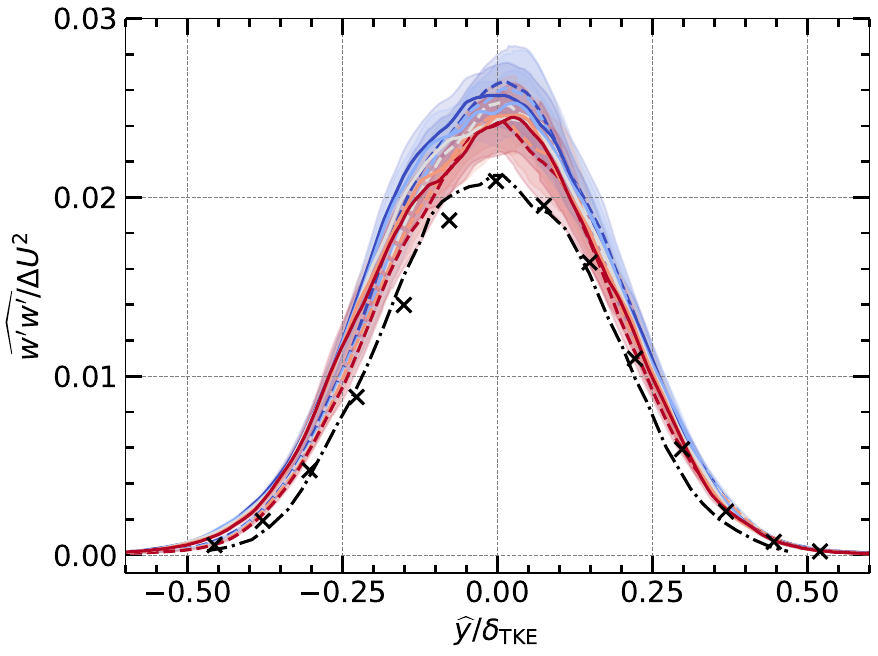}
  \end{subfigure}
  \hfill
  \begin{subfigure}[t]{0.49\columnwidth}
      \centering
      \includegraphics[width=\linewidth]{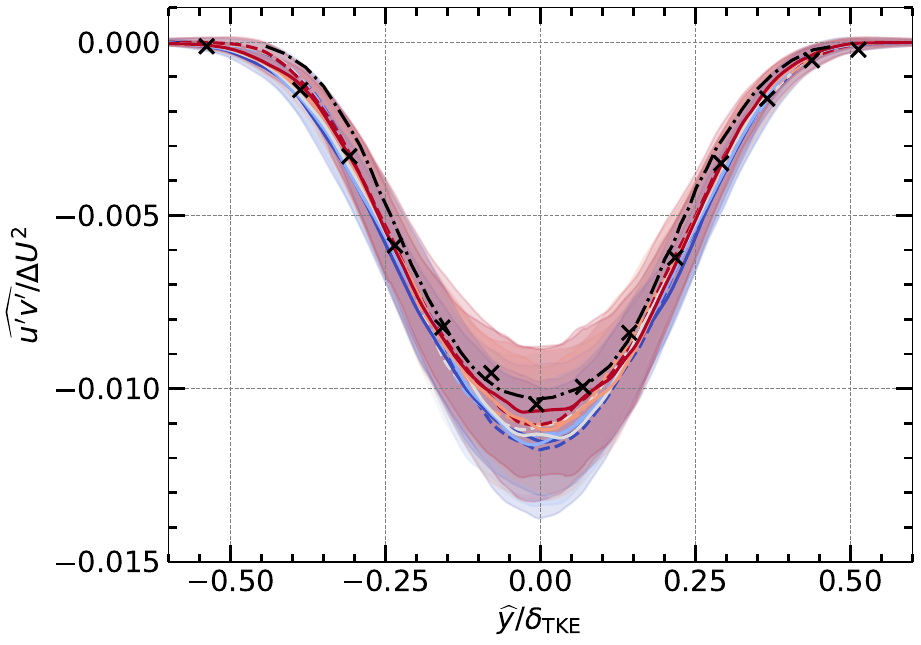}
  \end{subfigure}
 
 
  \begin{subfigure}[t]{0.49\columnwidth}
      \centering
      \includegraphics[width=\linewidth]{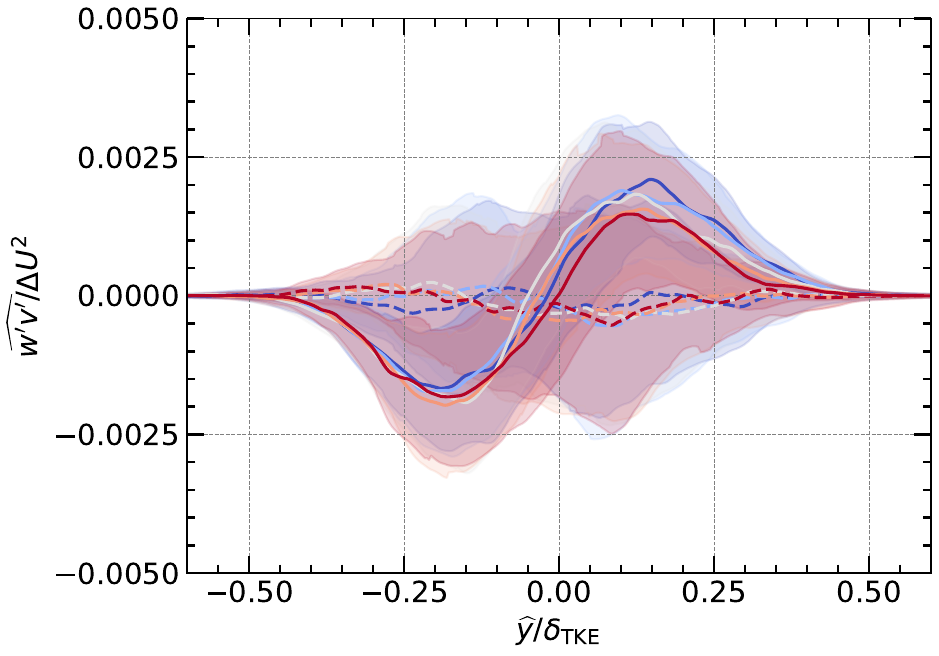}
  \end{subfigure}
  \hfill
  \begin{subfigure}[t]{0.49\columnwidth}
      \centering
      \includegraphics[width=\linewidth]{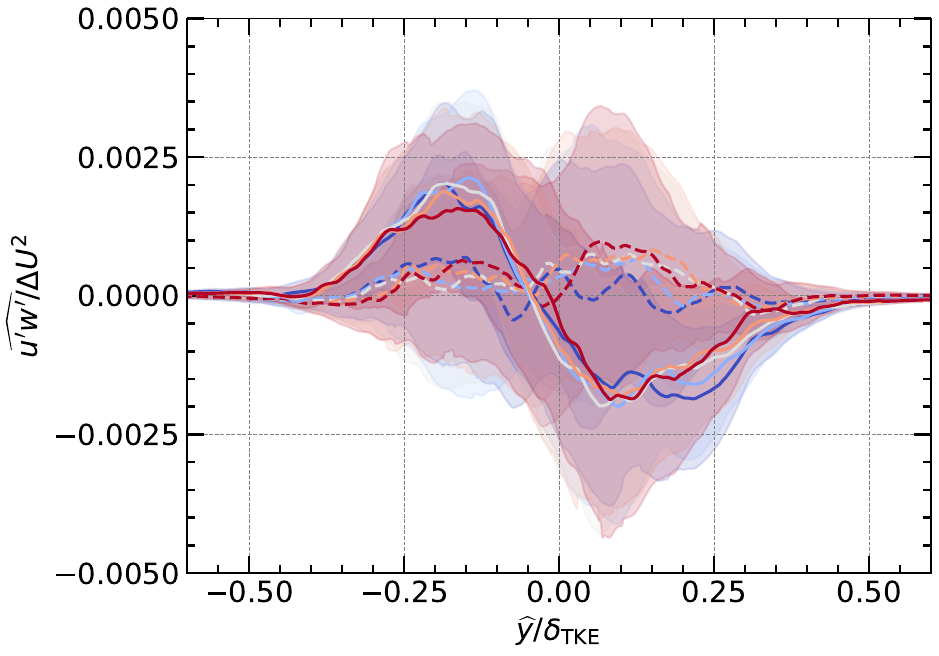}
  \end{subfigure}

  \caption{Reynolds stress tensor components in the rotated
  $(\boldvec{m},\boldvec{y},\boldvec{n})$ frame for the planar (dashed lines) and skewed (solid lines) shear layers
  during the middle and late times for the skewed and planar shear layers, respectively.
  The shaded areas represent the $95\%$ confidence intervals.
  Digitized data of \citet{Bell_AIAA_1990} (black `$\times$' symbols) and \citet{Rogers_PoF_1994} (black dash-dot line) 
  are overlaid for comparison in the top and middle rows.}
  \label{ReStressProfilesIntermediate}
\end{figure}

The evolution of $\widehat{u_i^\prime u_j^\prime}$ at later times is shown in figure~\ref{ReStressProfilesIntermediate}.
Note that the planar and skewed shear layers are in different flow regimes at this stage: the planar case is in the late times regime, the skewed case is in the middle times regime with a finite spanwise jet $\widehat{w}$.
Despite the presence of the spanwise jet, the normal stresses and $\widehat{u^\prime v^\prime}$ 
at different times agree rather well between the two cases (within the $95\%$ confidence interval), despite not quite having reached a self-similar state (there is still evidence of weak decay in time).
These results broadly agree with the digitized data of \citet{Bell_AIAA_1990} (black `$\times$' symbols) and \citet{Rogers_PoF_1994} (black dash-dot line)
for canonical shear layers. Overall, this observation strengthens the argument 
that the skewed shear layer evolution is only weakly linked to the decaying spanwise jet $\widehat{w}$. 

\begin{figure}
  \centering
  \includegraphics[width=0.6\columnwidth]{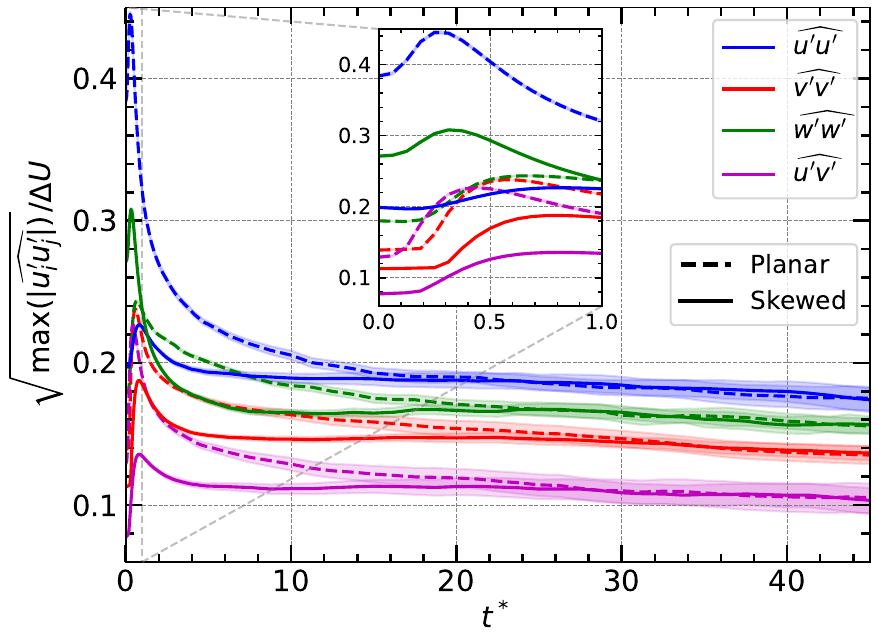}
  \caption{Evolution of the absolute maxima of Reynolds stress components (represented by different colors)
  for the planar (dashed lines) and skewed (solid lines) shear layers. The inset shows a zoomed view 
  of the results over $t^* \in [0,1]$, and the shaded areas represent the $95\%$ confidence intervals.}
\label{maxReStress}
\end{figure}

Unlike for $\widehat{u}$, these stress components have not quite reached a self-similar state and continue to decay monotonically 
throughout the entire observable time window.
This is highlighted by the maxima of the Reynolds stress components in figure~\ref{maxReStress}.
\citet{Pirozzoli_JFM_2015} also reported similar lack of saturation in Reynolds stresses for spatially evolving compressible shear layers in their 
computational domain. Based on the observations of \citet{Bell_AIAA_1990}, we expect Reynolds stresses to saturate at 
$t^* \sim \mathcal{O}(100)$, which is beyond the scope of current simulations due to domain size limitations. 

\begin{figure}
  \centering
  \begin{subfigure}[t]{0.49\columnwidth}
      \centering
      \includegraphics[width=\linewidth]{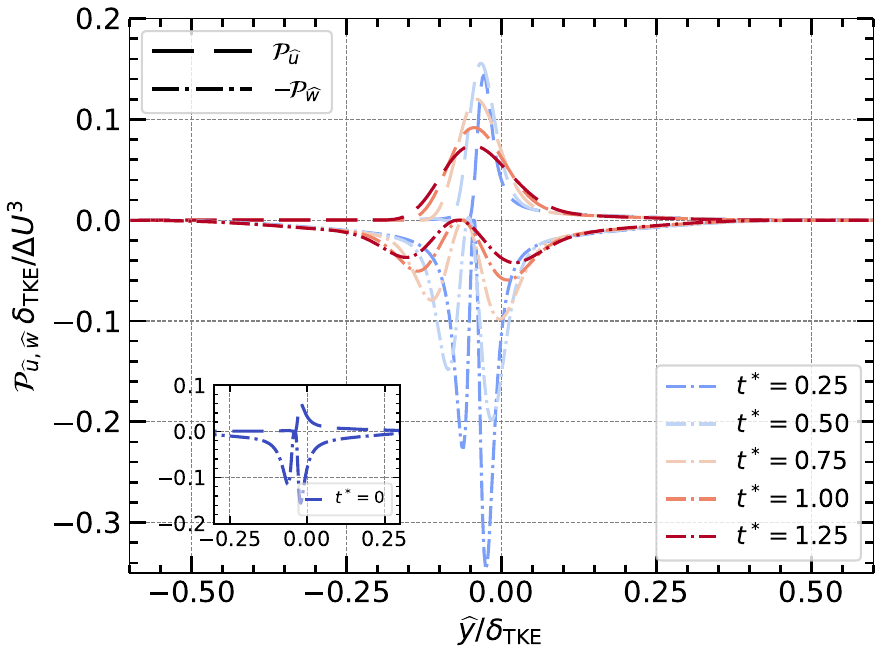}
      \caption{Early times.}
  \end{subfigure}
  \hfill
  \begin{subfigure}[t]{0.49\columnwidth}
      \centering
      \includegraphics[width=\linewidth]{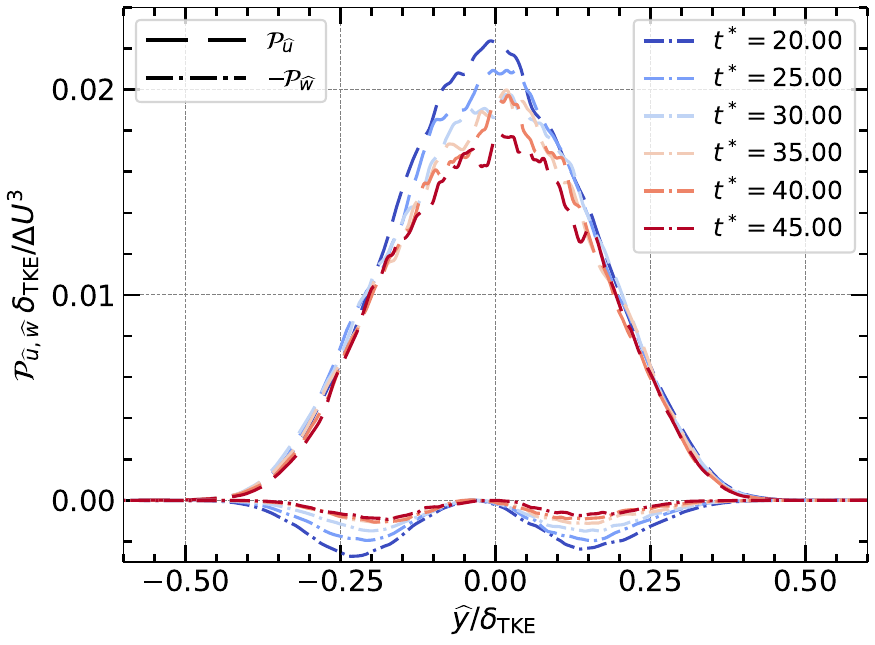}
      \caption{Middle times.}
  \end{subfigure} 
  \caption{Turbulence kinetic energy production components $\mathcal{P}_{\widehat{u}}$ (large dashed lines)
  and $\mathcal{P}_{\widehat{w}}$ (dash-dot lines; shown with a negative sign for clarity) for the skewed shear layer (see~\eqref{tkeProdEq}). The inset plot in the left figure 
  shows the initial condition at $t^*=0$.}
  \label{skewedTKEProdComp}
\end{figure}

\begin{figure}
  \centering
  \includegraphics[width=0.6\columnwidth]{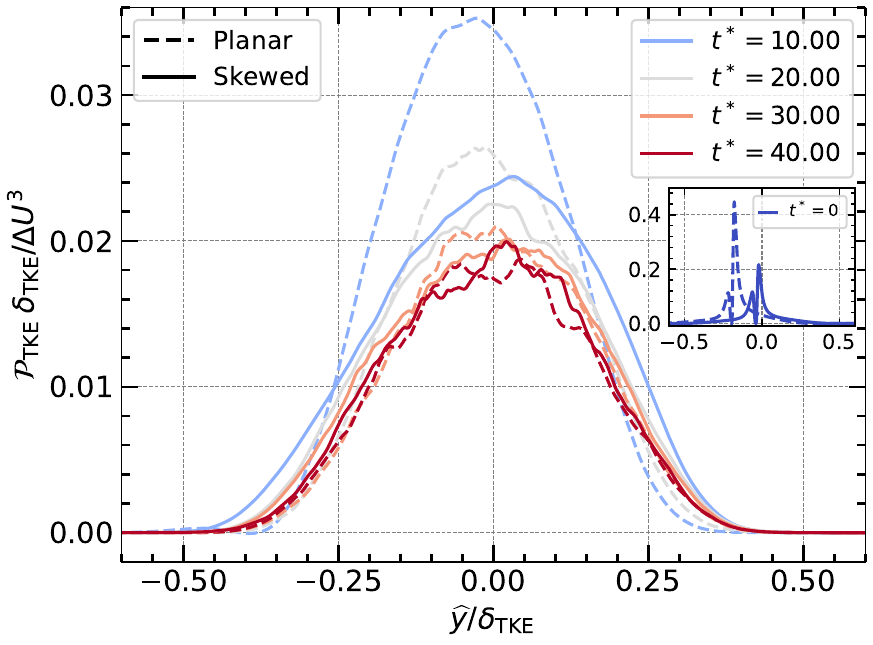}
  \caption{Turbulence kinetic energy production $\mathcal{P}_\text{TKE}$
  for the planar (dashed lines) and skewed (solid lines) shear layers.
  The inset shows the initial condition at $t^*=0$.}
  \label{tkeProd}
\end{figure}

The remaining shear stress components $\widehat{w^\prime v^\prime}$ and $\widehat{u^\prime w^\prime}$ 
in figure~\ref{ReStressProfilesIntermediate}
for the skewed shear layer are about an order of magnitude smaller than $\widehat{u^\prime v^\prime}$.
This is consistent with the reduced Reynolds stress production associated with the decaying $\widehat{w}$.
In addition, these components are not antisymmetric about $\widehat{y}=0$ (the shear layer symmetry plane), and instead, they gradually shift towards it.
This misalignment in turn affects the turbulence kinetic energy production
\begin{equation}
  \mathcal{P}_\text{TKE} = 
  \underbrace{\left(-\widehat{u^\prime v^\prime}\frac{\partial \widehat{u}}{\partial y}\right)}
  _{\substack{\mathcal{P}_{\widehat{u}}}}
  +
  \underbrace{\left(-\widehat{w^\prime v^\prime}\frac{\partial \widehat{w}}{\partial y}\right)}
  _{\substack{\mathcal{P}_{\widehat{w}}}},
  \label{tkeProdEq}
\end{equation}
with the different contributions shown in figure~\ref{skewedTKEProdComp}.
The production due to $\mathcal{P}_{\widehat{u}}$ is expected to be symmetric around $\widehat{y}=0$ and reach a self-similar state at late times; the former is true here but the latter has not quite been achieved by the end of the simulation.
The production due to the spanwise jet $\mathcal{P}_{\widehat{w}}$ is much smaller and decays in time, and shows the double-peak shape expected of a planar jet. It is quite clearly not symmetric around $\widehat{y}=0$.
Together, these differences in the early and middle times result in $\mathcal{P}_\text{TKE}$ for the skewed shear layer 
having a relatively ``fuller'' profile near the shear layer edges than the planar one (where $\mathcal{P}_{\widehat{w}}=0$), as shown in figure~\ref{tkeProd}. 

\section{Impact of skew on coherent structures}
The discussion so far has focused on the effects of three-component flow in a skewed shear layer on the mean flow statistics.
We now examine their impact on turbulence structures by analyzing the instantaneous flow data.

\subsection{Pressure rollers}
\begin{figure}
\noindent 
  \begin{subfigure}{0.32\columnwidth}
      \centering
      \includegraphics[width=\linewidth]{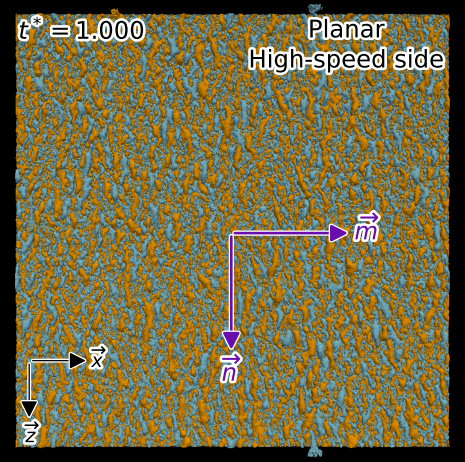}
  \end{subfigure}
  \hfill
  \begin{subfigure}{0.32\columnwidth}
      \centering
      \includegraphics[width=\linewidth]{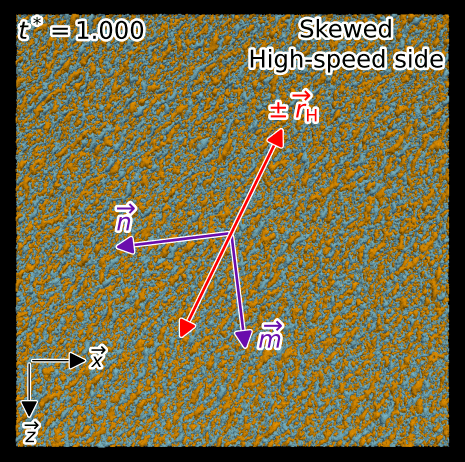}
  \end{subfigure}
  \hfill
  \begin{subfigure}{0.32\columnwidth}
      \centering
      \includegraphics[width=\linewidth]{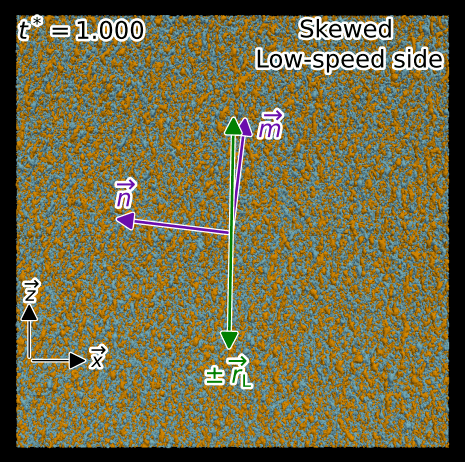}
  \end{subfigure}

  \vspace{0.1truein}

  \begin{subfigure}{0.32\columnwidth}
      \centering
      \includegraphics[width=\linewidth]{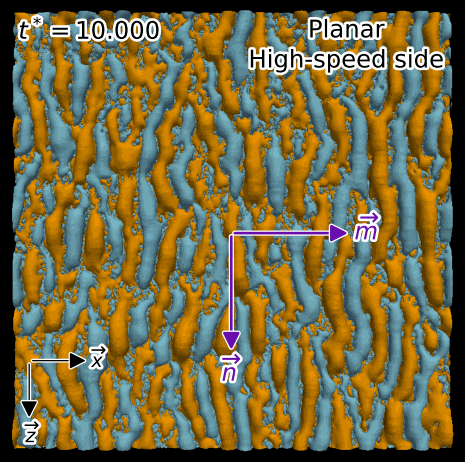}
  \end{subfigure}
  \hfill
  \begin{subfigure}{0.32\columnwidth}
      \centering
      \includegraphics[width=\linewidth]{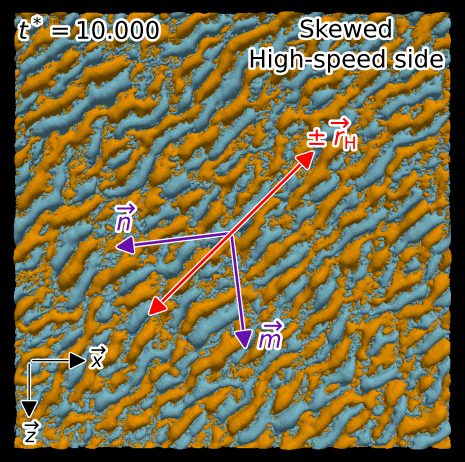}
  \end{subfigure}
  \hfill
  \begin{subfigure}[t]{0.32\columnwidth}
      \centering
      \includegraphics[width=\linewidth]{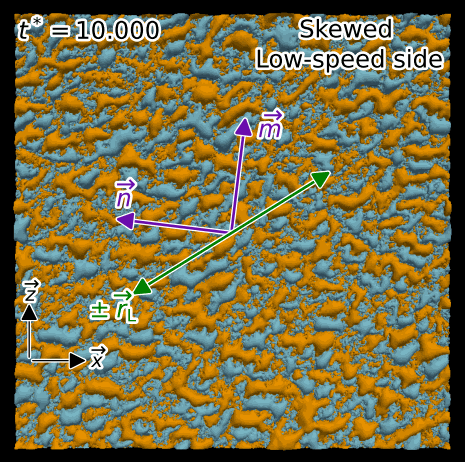}
  \end{subfigure}

  \vspace{0.1truein}

  \begin{subfigure}{0.32\columnwidth}
      \centering
      \includegraphics[width=\linewidth]{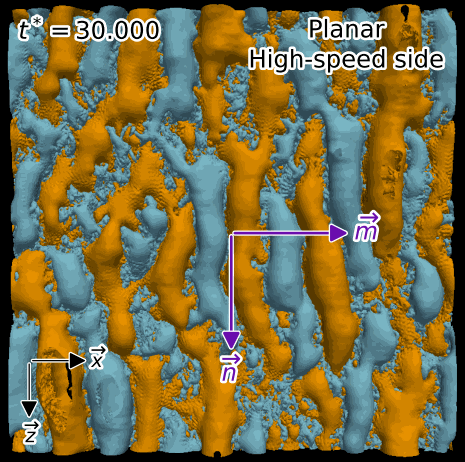}
  \end{subfigure}
  \hfill
  \begin{subfigure}{0.32\columnwidth}
      \centering
      \includegraphics[width=\linewidth]{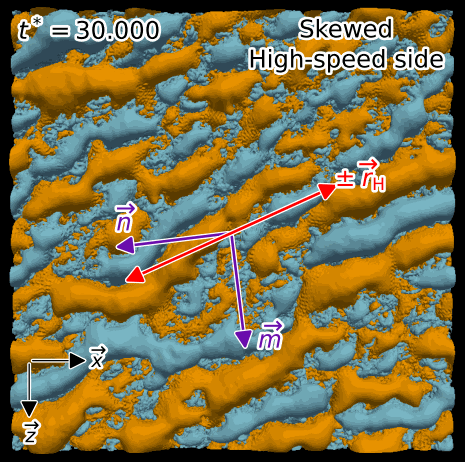}
  \end{subfigure}
  \hfill
  \begin{subfigure}{0.32\columnwidth}
      \centering
      \includegraphics[width=\linewidth]{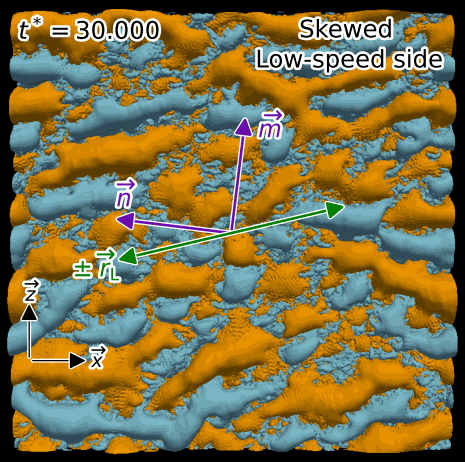}
  \end{subfigure}

  \caption{Isosurfaces of pressure fluctuations at three different times, with the lab frame $(\boldvec{x}, \boldvec{z})$
  and rotated mean shear frame $(\boldvec{m}, \boldvec{n})$ for the planar (left column) and skewed (middle and right columns)
  shear layers. The estimated orientation of the rollers for the skewed shear layer is overlaid as $\pm \boldvec{r}$.
  The visualizations can also be viewed as animations in Movie1.mp4 and Movie2.mp4, accessible at journals.cambridge.org.}
  \label{combinedPressureRollers}
\end{figure}

Previous studies on planar shear layers have highlighted the growth of Kelvin-Helmholtz instabilities 
into coherent spanwise vortices, commonly known as rollers in the literature \citep{Brown_JFM_1974,Rogers_PoF_1994}.
These rollers are visualized in the current study by plotting isosurfaces of pressure fluctuations at different time instances
in figure~\ref{combinedPressureRollers}. The rollers are similar when viewed from the high- and low-speed sides for the planar case, 
and hence only the former view is shown in the left column. In contrast, the coherent structures have changing orientations 
throughout the shear layer for the skewed case, and thus both views are shown in the middle and right columns.
The evolution of rollers for the two cases can also be viewed in animations Movie1.mp4 and Movie2.mp4 in the Supplementary Material.

\begin{figure}
  \centering
  \includegraphics[width=0.6\columnwidth]{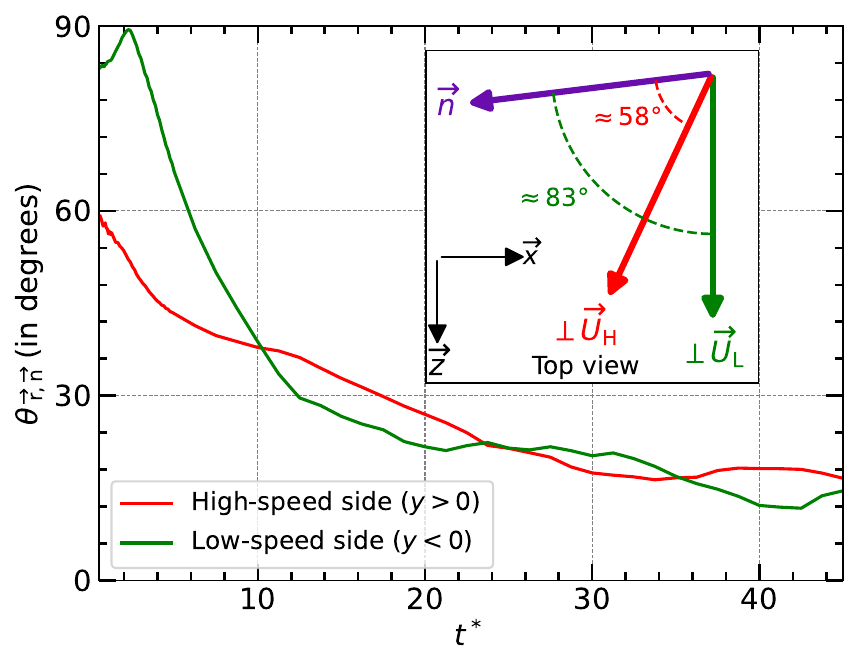}
  \caption{Evolution of the locally averaged misalignment $\theta_{\boldvec{r},\boldvec{n}}$.
  The inset shows directions orthogonal to $\boldvec{U}_\text{H}$ and $\boldvec{U}_\text{L}$, and their relative angle with the $\boldvec{n}$ direction 
  in the $\boldvec{x}-\boldvec{z}$ plane.}
\label{meanAngleRollers}
\end{figure}

The approximate orientation of these coherent structures is extracted by finding the eigenvalues $\lambda_\text{min/max}$ and eigenvectors $\boldvec{\nu}_\text{min/max}$ of the covariance matrix of the in-plane pressure gradient
\begin{equation}
\setlength{\arraycolsep}{3pt}
\renewcommand{\arraystretch}{1.5}
\mathcal{J} = \left(
\begin{array}{cc}
  \widebar{p^\prime_x p^\prime_x} & \widebar{p^\prime_x p^\prime_z}\\
  \widebar{p^\prime_x p^\prime_z} & \widebar{p^\prime_z p^\prime_z}\\ 
\end{array}  \right).
\label{pStructTensor}
\end{equation}
The eigenvector $\boldvec{\nu}_\text{min}$ should approximate the orientation of the rollers.
We anticipate that the rollers should align with the $\pm \boldvec{n}$ direction during late times, and thus define the misalignment as
\begin{equation}\label{rollerAngleEq}
\theta_{\vmin,\boldvec{n}} = \cos^{-1}\left(\left|\boldvec{v}_\text{min}\cdot\boldvec{n}\right|\right)
.
\end{equation}
This angle varies across the shear layer; we extract representative values by averaging it over
$|\widehat{y}/\deltaTKE| \in [0.45,0.5]$.
These mean angles $\theta_{\boldvec{r},\boldvec{n}}$ are visualized with arrows $\pm \boldvec{r}$ in figure~\ref{combinedPressureRollers}, and their evolution in time is shown in figure~\ref{meanAngleRollers}.
The initial misalignment angles are $\approx 58^\circ$ and $\approx 83^\circ$ for the high- and low-speed sides, respectively.
These values are consistent with the relative angles between $\boldvec{n}$ and the directions
orthogonal to $\boldvec{U}_\text{H}$ and $\boldvec{U}_\text{L}$ along which one would expect rollers in a ``shear layer'' between either side and stagnant flow ($\approx$ the wake in the initial condition).
At late times we would expect these angles to approach zero, when the rollers have become fully aligned with $\boldvec{n}$.
The finite values even at the end of the simulation again indicate that the skewed shear layer has 
not reached its late times state.

\subsection{Transverse correlations}
The preceding analysis highlighted the effect of skew on the organized large-scale motions across the shear layer.
We now assess the effect on the local coherence of the turbulent structures. To this end, we first compute
the two-point correlation of the transverse velocity fluctuations $v^\prime$ between any two locations $y$ and $y_1$ as
\begin{equation}\label{autoCorrEqn}
    A_{v^\prime v^\prime} \left(t^*,y,y_1\right) = \dfrac{\widebar{v^\prime(t^*,y)v^\prime(t^*,y_1)}}{\left[\widebar{v^{\prime} (t^*,y)^2}\;\widebar{v^{\prime} (t^*,y_1)^2}\right]^{1/2}},
\end{equation}
similar to \citet{Matsuno_JFM_2021}. Note that the choice of computing the correlation of $v^\prime$ is motivated by their larger sensitivity to cross-stream
mixing across the shear layer thickness. The correlation function is then integrated along $y_1$ to obtain an effective integral length scale, 
\begin{equation}\label{autoCorrLength}
    \mathcal{L}_{v^\prime v^\prime} \left(t^*,y \right)= \int_{y_l}^{y_u} A_{v^\prime v^\prime} \left(t^*,y,y_1\right) \; dy_1,
\end{equation}
quantifying the transverse extent over which the fluctuations remain correlated about the reference location $y$.
The 
integration bounds $y_l$ and $y_u$ are defined as the smaller of the distance $\pm \deltaTKE$ about the instantaneous shear layer centerline 
$\widehat{y}=0$, or the first zero-crossing location for $A_{v^\prime v^\prime}$.

\begin{figure}
  \centering
  \includegraphics[width=0.6\columnwidth]{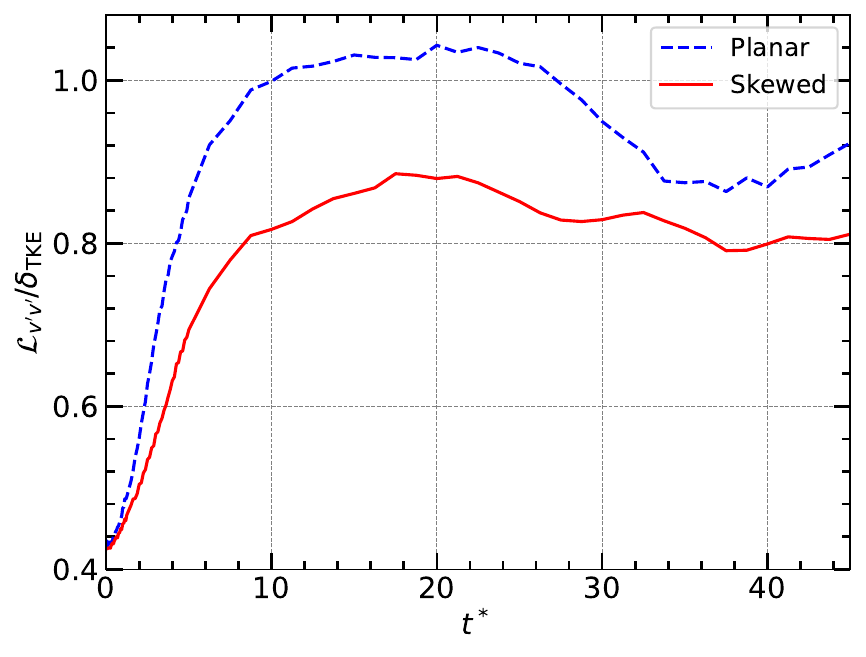}
  \caption{Evolution of the normalized integral length scale $\mathcal{L}_{v^\prime v^\prime}/\deltaTKE$ for the planar (dashed line) 
  and skewed (solid line) shear layers evaluated near their edges (mean of results at $\widehat{y}/\deltaTKE = \pm 0.4$).}
\label{autoCorrLengthScaleVsTime}
\end{figure}

We compare $\mathcal{L}_{v^\prime v^\prime}/\deltaTKE$ near the shear layer edges, where the differences are most pronounced. To this end, we compute
the mean of $\mathcal{L}_{v^\prime v^\prime}/\deltaTKE$ at $\widehat{y}/\deltaTKE = \pm 0.4$, with the results shown in figure~\ref{autoCorrLengthScaleVsTime}.
For both cases, the coherence grows monotonically over $0 \le t^* \lesssim 14$. The planar shear layer shows stronger growth and hence transiently saturates
at larger $\mathcal{L}_{v^\prime v^\prime}/\deltaTKE$ as compared to the skewed shear layer. Following this, coherence decays and approaches saturation at 
comparable values for both cases, indicating convergence towards similar late times states. Therefore, this result indicates that transient three-component
flow in the skewed shear layer leads to reduced coherence in turbulent structures, and the effect diminishes with the flow approaching the two-component
planar state further downstream.

\section{Numerical experiments using fictitious cases}\label{secFictCases}
The analysis so far has highlighted key three-component effects in skewed shear layers.
However, since both mean flow and turbulence have a three-component nature (at least initially),
it is unclear how they contribute to these observed effects. To address this question, we set up and study 
a set of fictitious test cases as controlled ``what if?'' type numerical experiments.

\subsection{Isolated misalignment in mean flow and turbulence}
We first examine if the observed three-component effects are primarily mediated through misalignment in 
the mean flow, the turbulence, or a combination of both. To this end, we isolate misalignment in mean
flow and turbulence by interchanging them between the planar and skewed cases as  
\begin{equation}
\begin{aligned}
    \phi_{\overline{\text{planar}},\text{skewed}^\prime} &= \widebar{\phi}_\text{planar} + \phi^\prime_\text{skewed},\\
    \phi_{\overline{\text{skewed}},\text{planar}^\prime} &= \widebar{\phi}_\text{skewed} + \phi^\prime_\text{planar},
\end{aligned}
\end{equation}
where $\phi$ represents the instantaneous flow state at $t^*=0$. 
In other words, the case `$\overline{\text{planar}},\text{skewed}^\prime$' has 
(at $t^*=0$) 
a non-skewed mean flow but
a turbulence field from the skewed case, and vice versa for the `$\overline{\text{skewed}},\text{planar}^\prime$'  case.

\begin{figure} 
  \begin{subfigure}[t]{0.5\columnwidth}
      \centering
      \includegraphics[width=\linewidth]{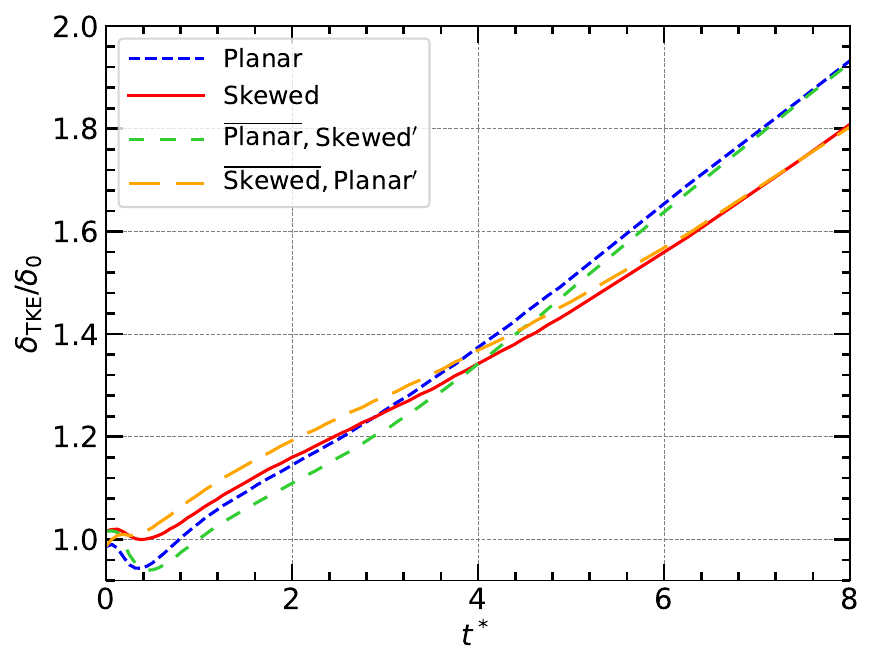}
  \end{subfigure}
  \hfill
  \begin{subfigure}[t]{0.5\columnwidth}
      \centering
      \includegraphics[width=1.05\linewidth]{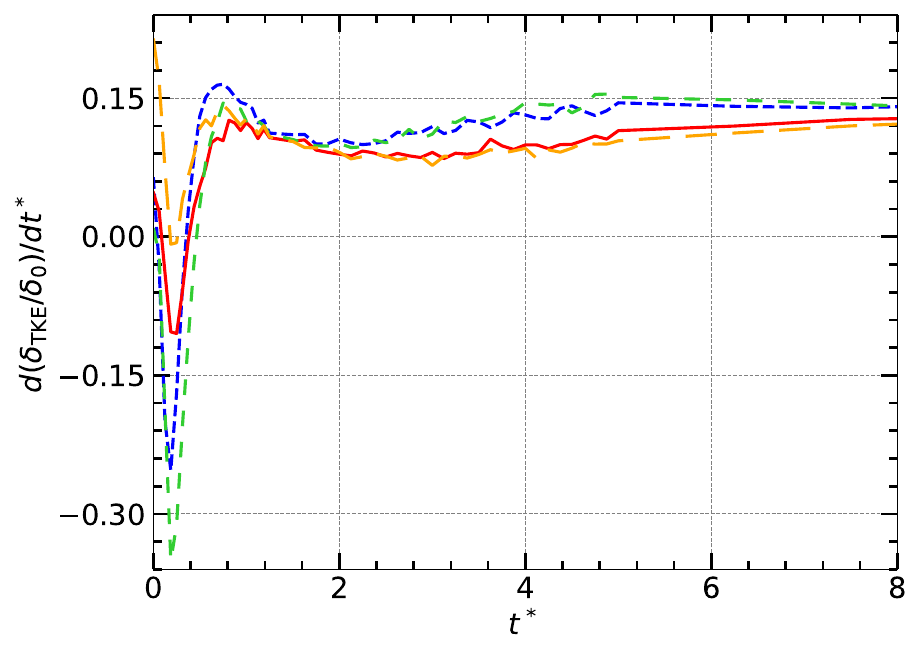}
  \end{subfigure} 
  \caption{Evolution of the shear layer thickness $\deltaTKE$ (left) and its growth rate (right) for the fictitious cases, combined with the planar 
  and skewed cases.}
  \label{fictSwappedDeltaTKE}
\end{figure}

We first compare the growth of $\deltaTKE$ in 
figure~\ref{fictSwappedDeltaTKE}. Since $\deltaTKE$ is computed using the turbulence kinetic energy profile, cases 
with the same initial turbulence field $\phi^\prime$ have identical values at $t^*=0$. The results indicate rapid 
readjustment in the fictitious cases over $0 < t^* \lesssim 0.5$, driven towards behavior consistent with the imposed mean flow. 
Consequently, cases with the same mean flow exhibit similar growth rates for $t^* \gtrsim 1$.

\begin{figure}
  \centering
  \begin{subfigure}[t]{0.49\columnwidth}
      \centering
      \includegraphics[width=\linewidth]{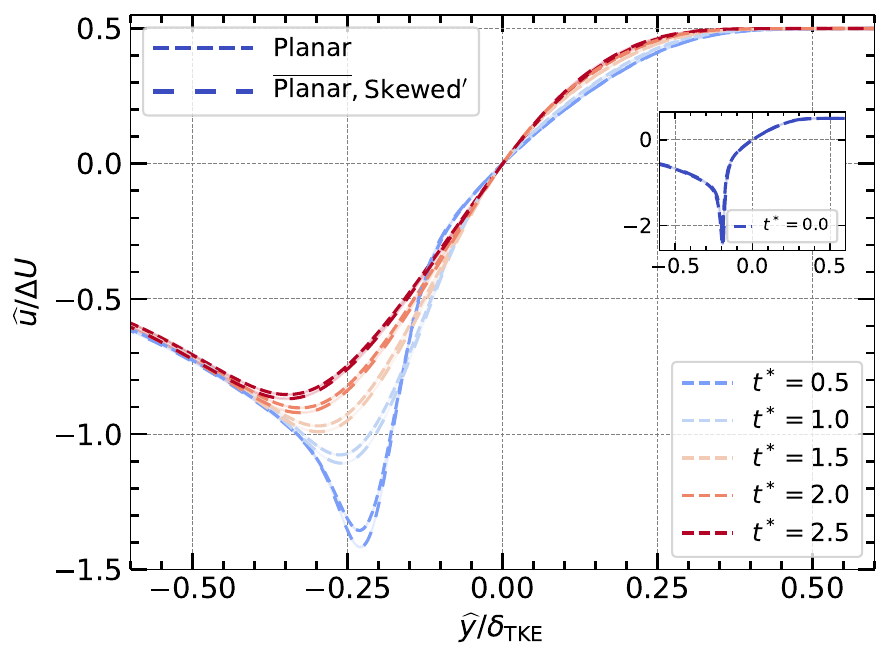}
  \end{subfigure}
  \hfill
  \begin{subfigure}[t]{0.49\columnwidth}
      \centering
      \includegraphics[width=\linewidth]{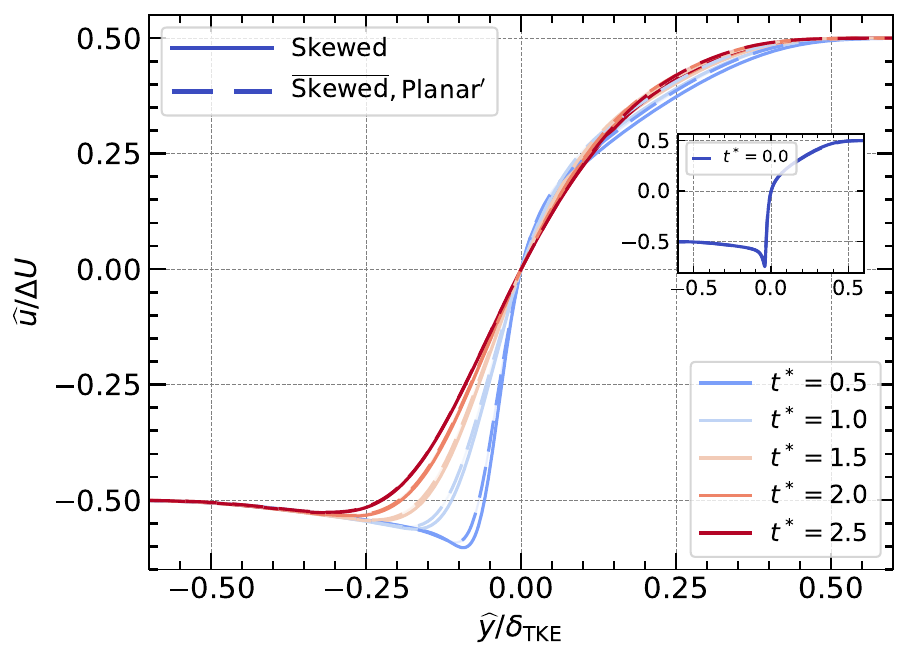}
  \end{subfigure} 

  \vspace{0.1truein}
 
  \begin{subfigure}[t]{0.49\columnwidth}
      \centering
      \includegraphics[width=\linewidth]{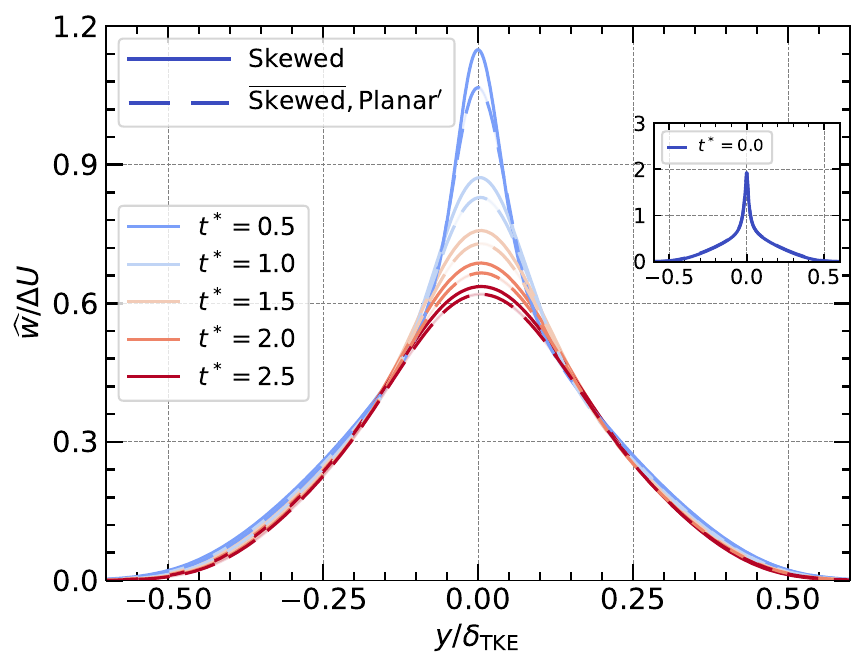}
  \end{subfigure} 
  \caption{Mean velocity profiles for the fictitious cases, compared against those of the planar and skewed shear layers.
  The insets show initial conditions for the respective cases (note that the two cases in each inset overlap).
  The shaded areas around the mean profiles represent the 95\% confidence intervals.}
  \label{fictSwappedVelocityProfiles}
\end{figure}

\begin{figure}
  \centering
  \begin{subfigure}[t]{0.49\columnwidth}
      \centering
      \includegraphics[width=\linewidth]{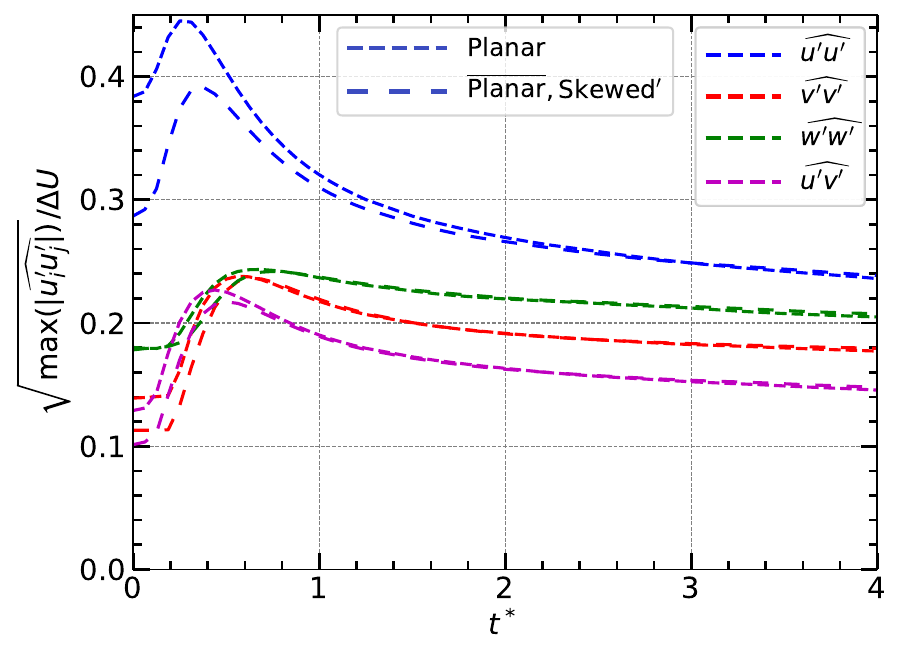}
  \end{subfigure}
  \hfill
  \begin{subfigure}[t]{0.49\columnwidth}
      \centering
      \includegraphics[width=\linewidth]{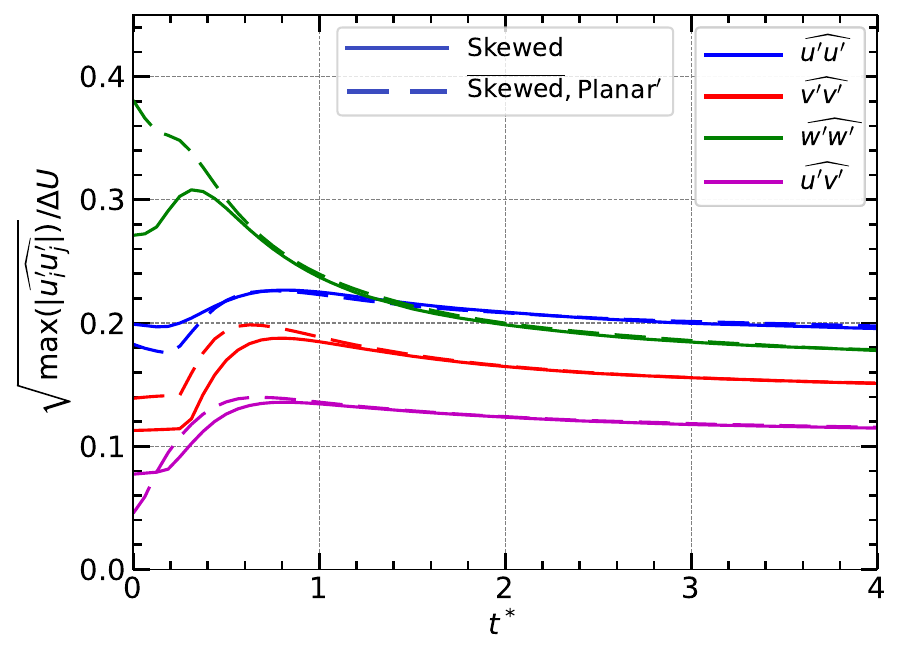}
  \end{subfigure} 

  \vspace{0.1truein}
 
  \begin{subfigure}[t]{0.49\columnwidth}
      \centering
      \includegraphics[width=\linewidth]{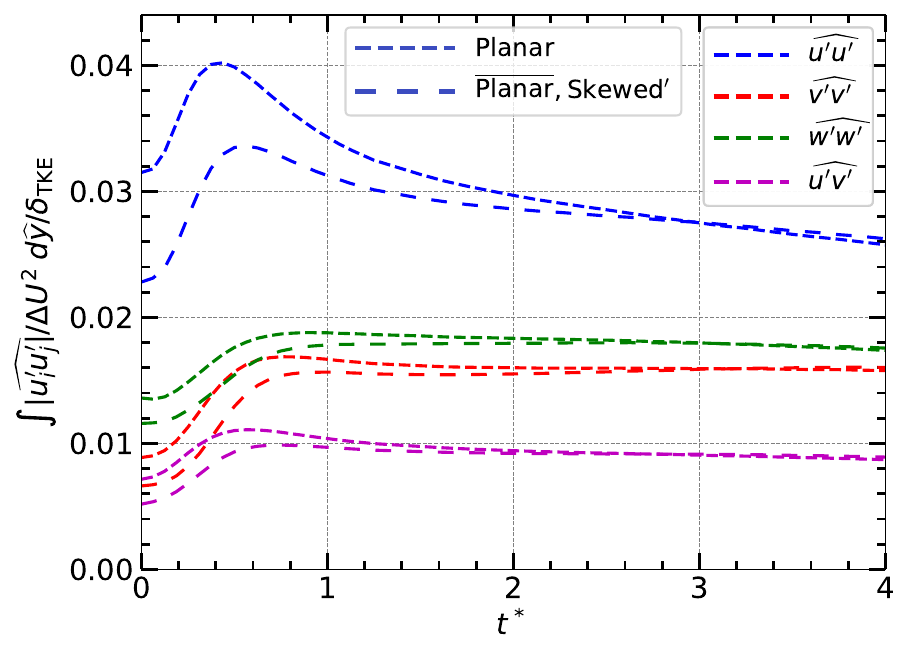}
  \end{subfigure}
  \hfill
  \begin{subfigure}[t]{0.49\columnwidth}
      \centering
      \includegraphics[width=\linewidth]{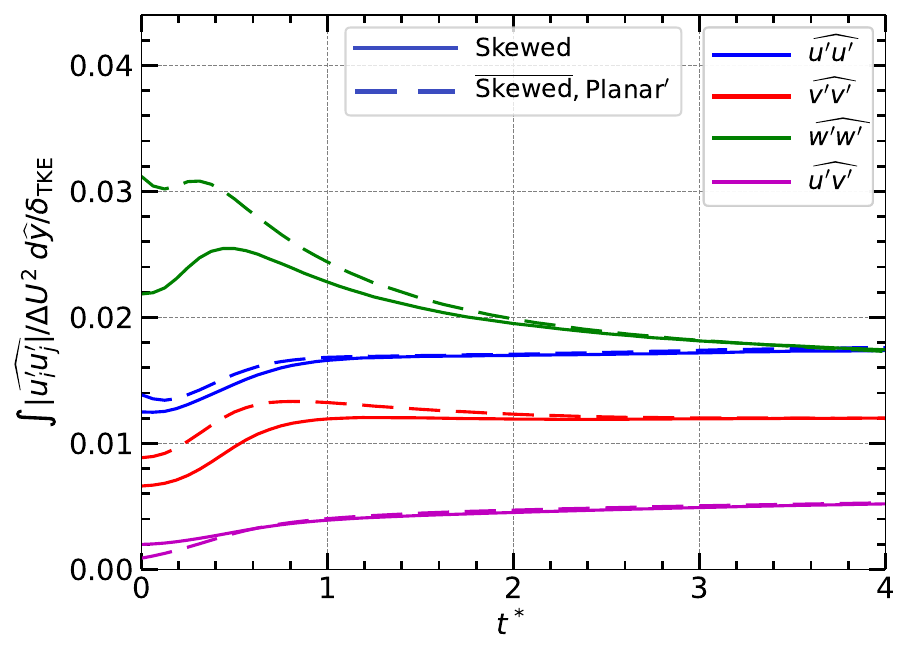}
  \end{subfigure} 
  \caption{Evolution of the maxima (top row) and integrated (bottom row) Reynolds stress tensor components (represented by different line colors) 
  in the $(\boldvec{m},\boldvec{y},\boldvec{n})$ frame for the fictitious cases, compared against the planar and skewed cases.}
  \label{fictSwappedReStressVsTime}
\end{figure}

This motivates the comparison of mean $\widehat{u}$ and $\widehat{w}$ profiles in figure~\ref{fictSwappedVelocityProfiles}.
The impact of the altered initial turbulence on the mean flow is minimal, with only minor differences observed 
for $0 < t^* \lesssim 2.5$. Specifically, cases with turbulence imposed from the planar shear layer 
(cases planar and $\overline{\text{skewed}},\text{planar}^\prime$) exhibit slightly faster decay in regions of large velocity gradients, 
after which the profiles overlap (selected results shown for brevity). This behavior can be attributed to larger initial Reynolds stresses at 
$t^*=0$ in the planar shear layer (by design, see figure~\ref{ReStressProfilesInit}), which leads to stronger turbulent diffusion in the 
early stage of evolution. However, the Reynolds stresses rapidly readjust with the imposed mean flow, as indicated by 
the evolution of their maxima and integrals across the shear layer in figure~\ref{fictSwappedReStressVsTime}. Consequently, the mismatch in
Reynolds stresses and the mean velocity profiles diminishes over similar time scales.

\begin{figure} 
  \begin{subfigure}[t]{0.5\columnwidth}
      \centering
      \includegraphics[width=\linewidth]{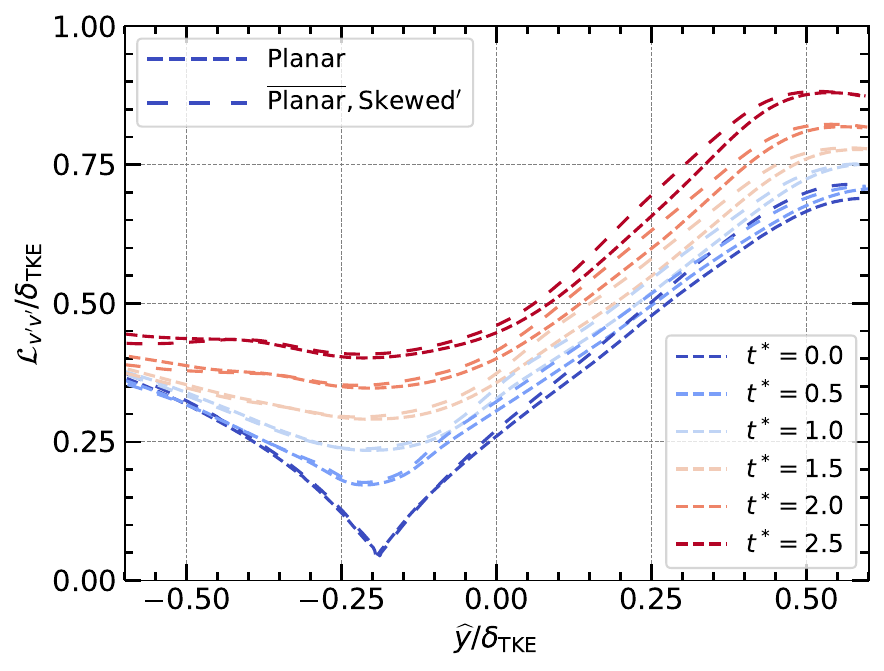}
  \end{subfigure}
  \hfill
  \begin{subfigure}[t]{0.5\columnwidth}
      \centering
      \includegraphics[width=\linewidth]{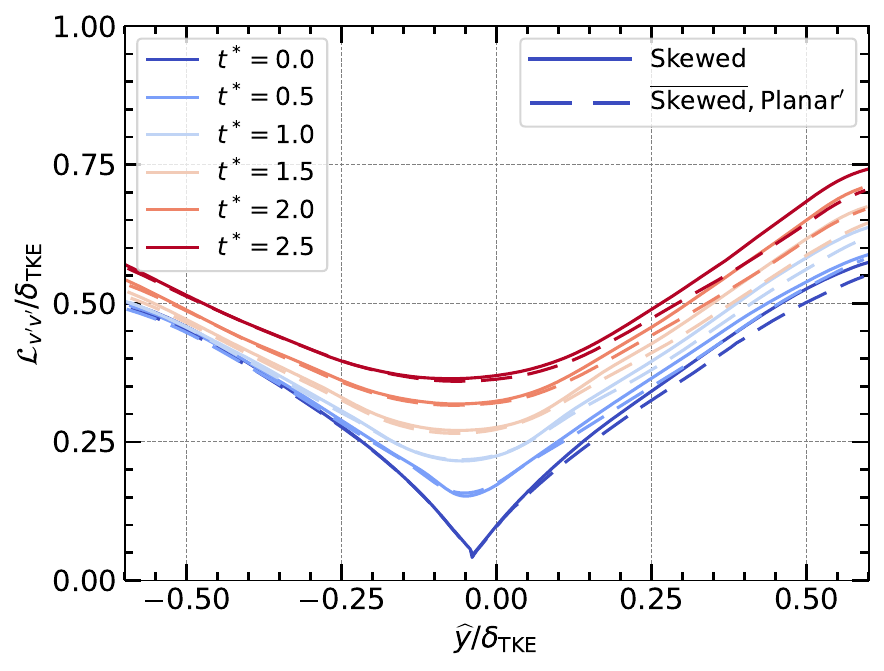}
  \end{subfigure} 
  \caption{Evolution of the normalized integral length scale $\mathcal{L}_{v^\prime v^\prime}/\deltaTKE$ for the 
  fictitious cases, compared against the planar and skewed cases.}
  \label{fictSwappedVCorrInt}
\end{figure}

In addition, we assess the effect of misalignment on transverse 
coherence by comparing the evolution of $\mathcal{L}_{v^\prime v^\prime}/\deltaTKE$ (see~\eqref{autoCorrLength})
in figure~\ref{fictSwappedVCorrInt}. The results indicate a similar trend, where misalignment in mean flow (cases skewed and $\overline{\text{skewed}},\text{planar}^\prime$, right subplot) 
leads to reduced coherence. Comparatively, the impact of misalignment in the initial turbulence field is minimal, and the small differences could be attributed 
to the early mismatch in $\deltaTKE$ (see figure~\ref{fictSwappedDeltaTKE}).
Together, these results indicate that 
the three-component flow behavior observed in skewed shear layers is primarily driven by misalignment in the mean flow.

\subsection{Planar shear layer + artificial spanwise jet}\label{secPlanarJet}
\begin{figure}
  \centering
  \includegraphics[width=0.6\textwidth]{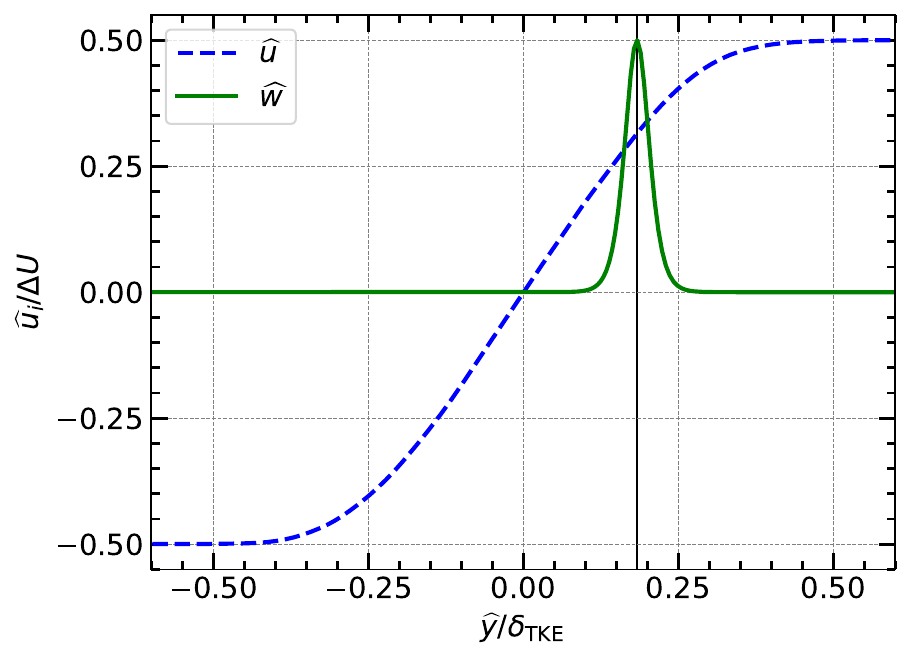}
  \caption{Mean velocity profiles for the artificial jet case at $t^*=20$,
  the time instance at which the artificial spanwise jet $\widehat{w}$ is superimposed onto the planar shear layer snapshot. The solid black
  line represents the centerline $\widehat{y}_{c}/\deltaTKE \approx 0.2$ for the artificial jet.}
\label{planarJetSetupFig}
\end{figure}

For the skewed shear layer, the self-similar evolution of $\widehat{u}$ alongside the planar jet-like decay of $\widehat{w}$
(see Figs.~\ref{uHatProfiles} and~\ref{wHatJetProfiles}) suggests that the flow in the $\boldvec{m}$ and $\boldvec{n}$ direction are 
only weakly coupled to one another. This observation is similar to the ``independence principle'' in laminar boundary layers, where the 
streamwise and spanwise flow evolution can be considered to be mutually decoupled from one another. However, in turbulent flows, the applicability
of the principle is only supported in limited cases \citep{Wygnanski_JFM_2011,Wygnanski_JA_2014} due to the coupled turbulent shear stresses 
appearing in the mean momentum equations. So instead, we assess the degree of independence
between $\widehat{u}$ and $\widehat{w}$ by constructing a fictitious test case (labeled as ``planar+jet''), 
where an artificial mean spanwise jet profile is superimposed onto an instantaneous snapshot of a planar shear layer. Here, we pick the shear 
layer snapshot at $t^*_\text{planar+jet} = 20$, such that $t^*_\text{endDeficit} < t^*_\text{planar+jet} < t^*_\text{endEarly}$. The superposed 
jet profile is defined using~\eqref{jetSelfSimilarEq}, with parameters $\widehat{w}_c/\Delta U = 0.5$, $y_{1/2}/\deltaTKE \approx 1/40$, and an 
offset centerline location $\widehat{y}_c/\deltaTKE \approx 0.2$. For reference, a comparison between the mean $\widehat{u}$ and $\widehat{w}$ 
profiles is shown in figure~\ref{planarJetSetupFig}.

\begin{figure} 
  \begin{subfigure}{0.5\columnwidth}
      \centering
      \includegraphics[width=\linewidth]{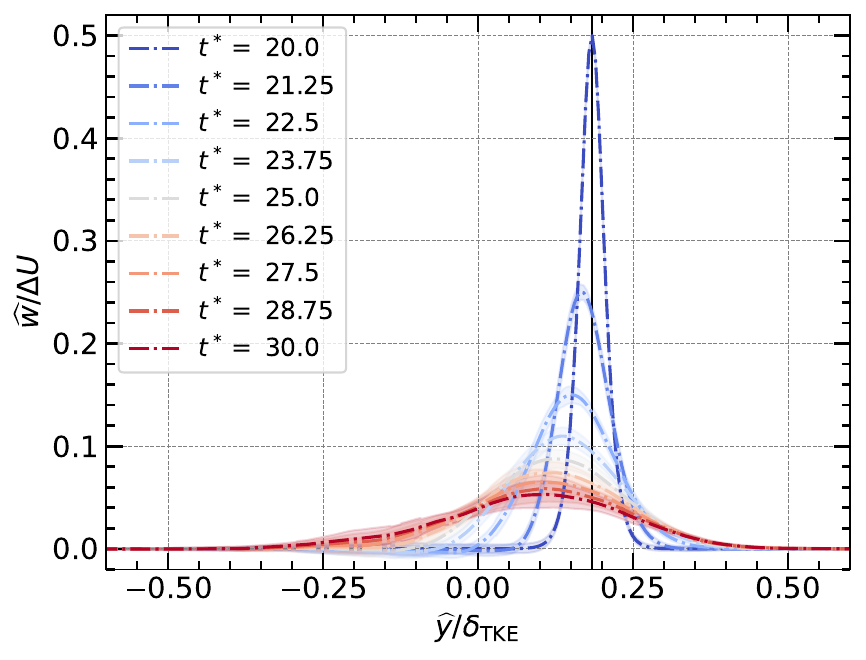}
      \caption{Evolution of artificial $\widehat{w}$.}
  \end{subfigure}
  \hfill
  \begin{subfigure}{0.5\columnwidth}
      \centering
      \includegraphics[width=1.075\linewidth]{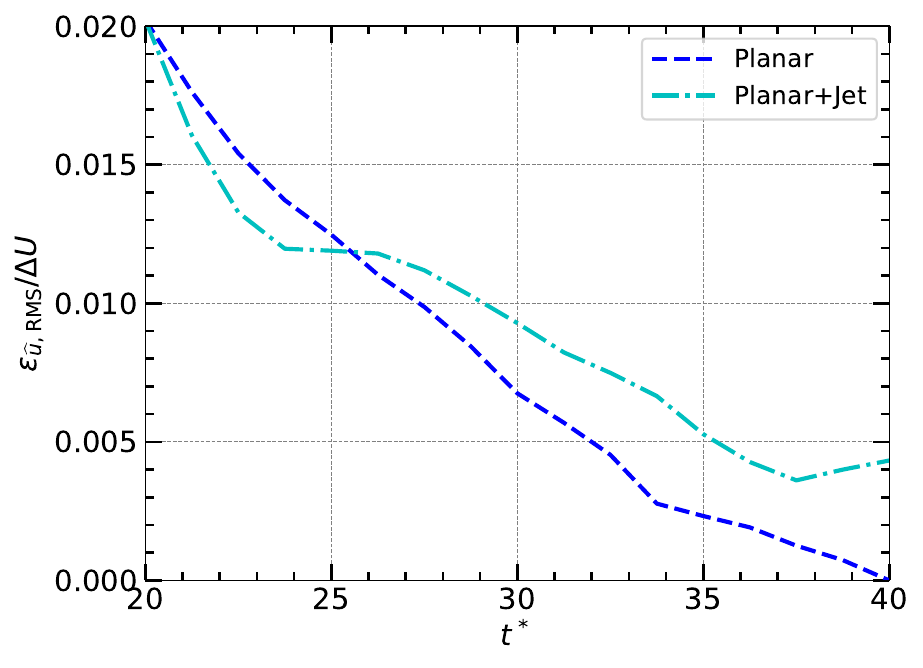}
      \caption{Decay of $\varepsilon_{\widehat{u},\mathrm{RMS}}(t^*)$.}
  \end{subfigure} 
  \caption{Assessment of the mean velocity profiles for the planar+jet case, starting at $t^*_\text{planar+jet} = 20$.
The right figure also includes the planar case from figure~\ref{rmsErrorUhat}.}
  \label{planarJetVelocityProfiles}
\end{figure}

We begin by examining the evolution of the mean velocity profiles in figure~\ref{planarJetVelocityProfiles}. The left subplot shows the rapid 
decay of the artificial $\widehat{w}$, with the jet centerline shifting towards the shear layer centerline $\widehat{y}=0$, consistent with restoration
of the late times symmetry of the shear layer. To quantify its impact on $\widehat{u}$, we plot $\varepsilon_{\widehat{u},\mathrm{RMS}}(t^*)$ 
(see~\eqref{rmsErrorUhatEq}, with $\widehat{u}_\mathrm{ref}$ taken from the planar case at $t^*=40$) in the right subplot. For comparison, we also overlay the corresponding results for the planar shear layer. Despite the 
artificial $\widehat{w}_c$ being initially comparable in magnitude to freestream $\widehat{u}$, the impact on the latter is minimal
with the error differing by $\lesssim 0.5\%$ of $\Delta U$.

\begin{figure}
\noindent 
  \begin{subfigure}{0.5\columnwidth}
      \centering
      \includegraphics[width=\linewidth]{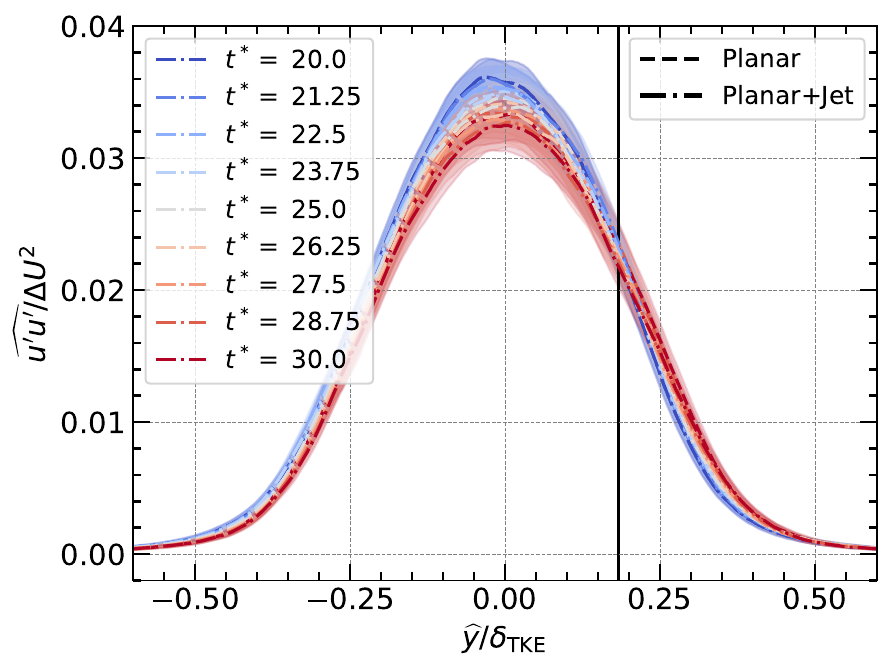}
  \end{subfigure}
  \hfill
  \begin{subfigure}{0.5\columnwidth}
      \centering
      \includegraphics[width=\linewidth]{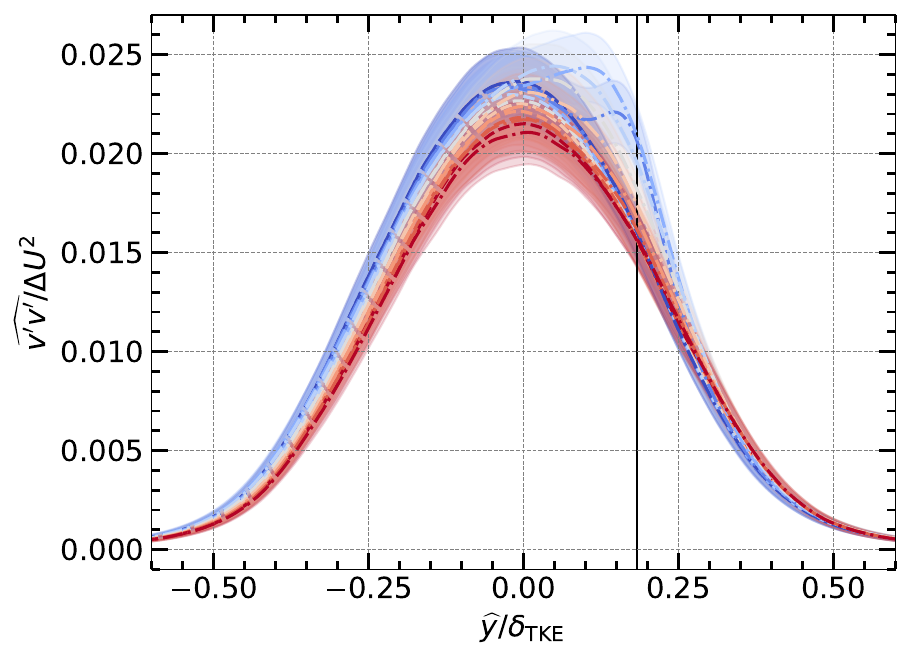}
  \end{subfigure}

  \vspace{0.1truein}

  \begin{subfigure}{0.5\columnwidth}
      \centering
      \includegraphics[width=\linewidth]{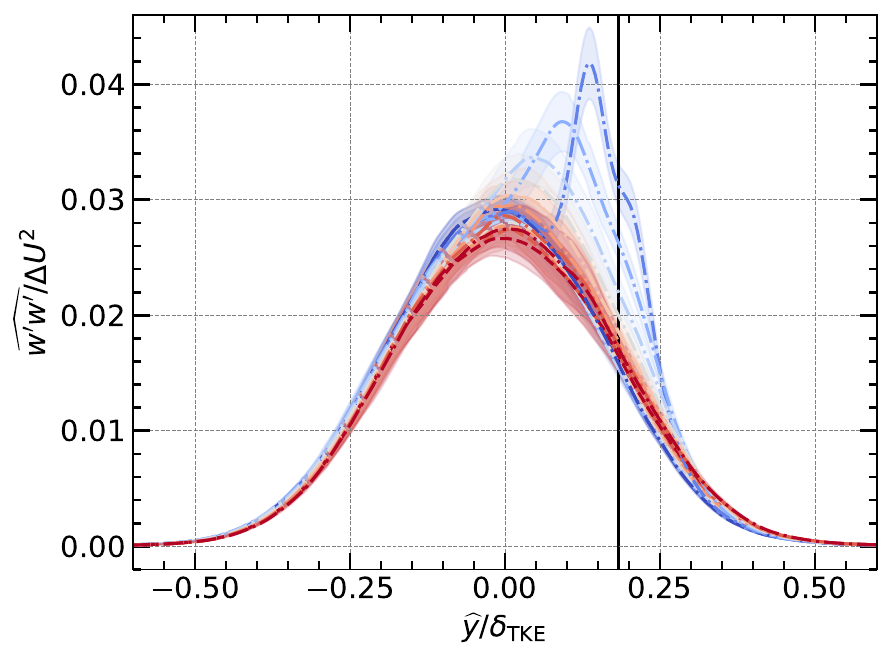}
  \end{subfigure}
  \hfill
  \begin{subfigure}{0.5\columnwidth}
      \centering
      \includegraphics[width=\linewidth]{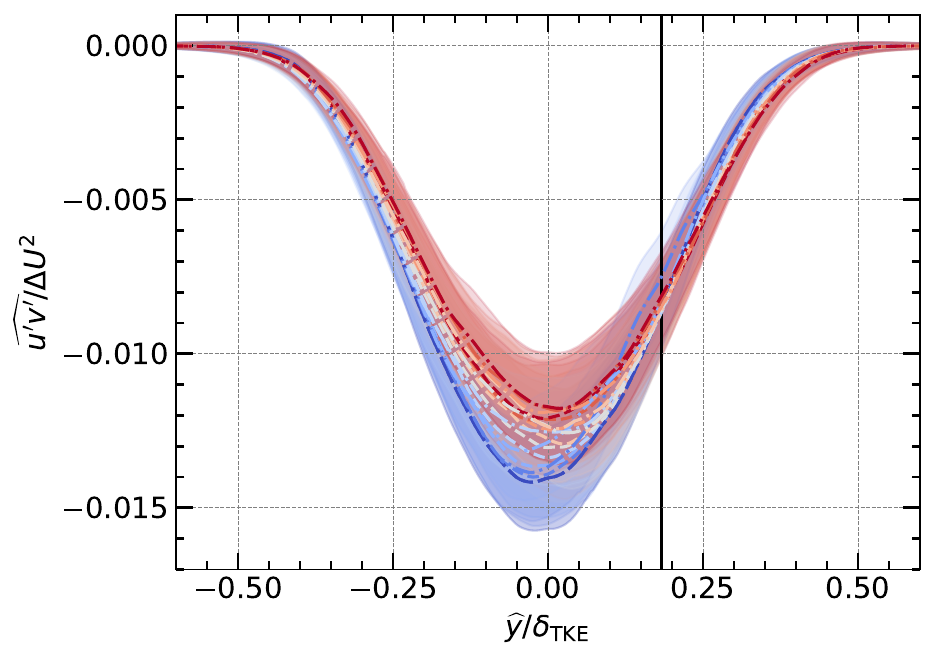}
  \end{subfigure}

  \vspace{0.1truein}

  \begin{subfigure}{0.5\columnwidth}
      \centering
      \includegraphics[width=\linewidth]{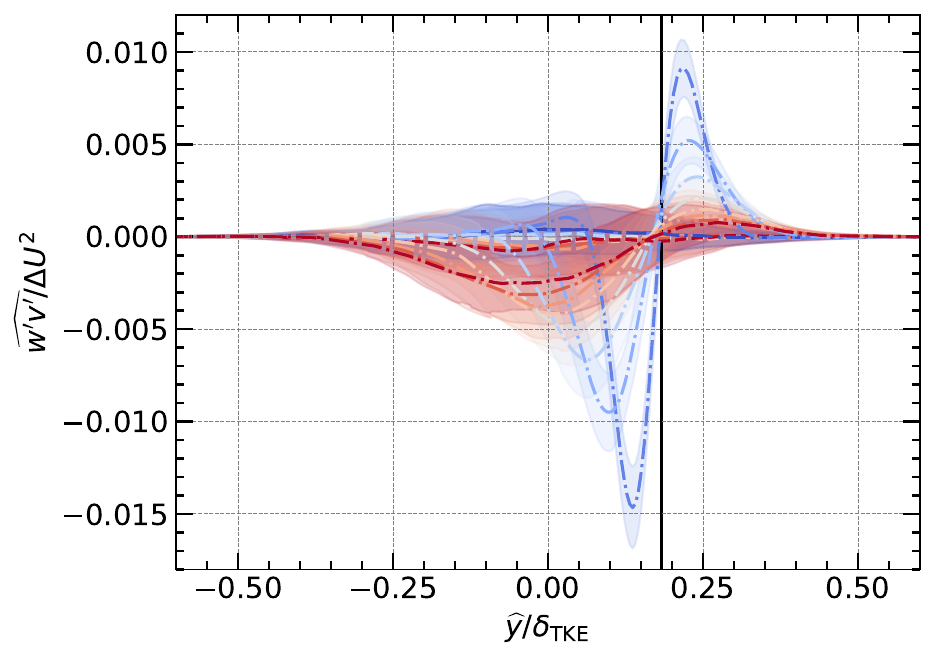}
  \end{subfigure}
  \hfill
  \begin{subfigure}{0.5\columnwidth}
      \centering
      \includegraphics[width=\linewidth]{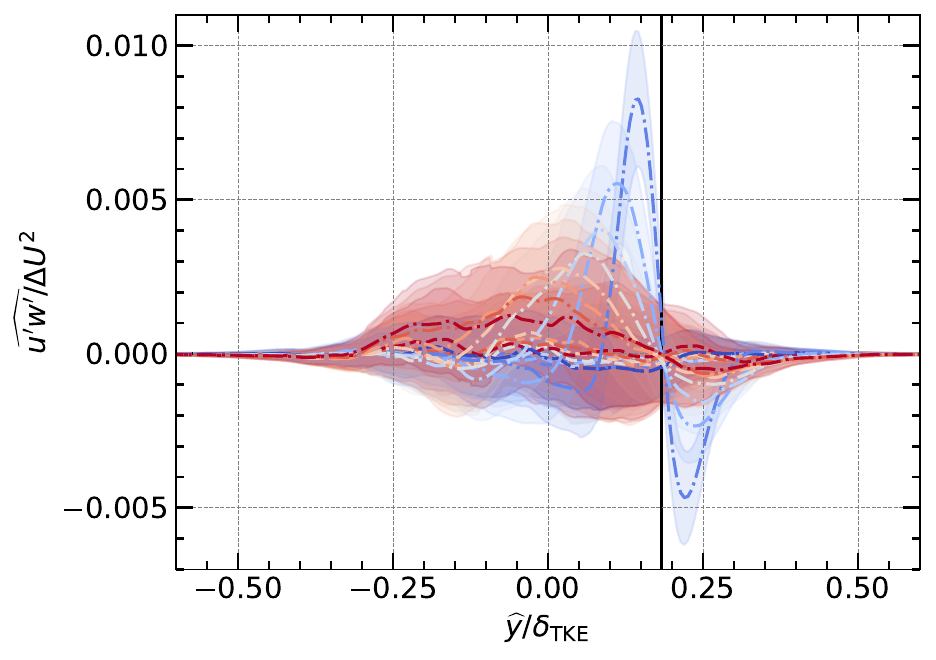}
  \end{subfigure}
  \caption{Reynolds stress profiles in the $(\boldvec{m},\boldvec{y},\boldvec{n})$
  frame for the planar and planar+jet cases.
  The shaded areas represent the $95\%$ confidence intervals.
  The black vertical line represents the 
  artificial jet centerline $\widehat{y}/\deltaTKE \approx 0.2$ for the planar+jet case.}
  \label{planarJetReStressProfiles}
\end{figure}

We next compare the evolution of the Reynolds stress tensor against the baseline planar shear layer in figure~\ref{planarJetReStressProfiles}.
The components $\widehat{w^\prime w^\prime}$, $\widehat{w^\prime v^\prime}$, and $\widehat{u^\prime w^\prime}$ are the most strongly affected, 
exhibiting a rapid increase in magnitude near the imposed jet centerline $\widehat{y}_c/\deltaTKE \approx 0.2$ (marked by the solid black line) due to increased 
Reynolds stress production.
The remaining components $\widehat{v^\prime v^\prime}$, $\widehat{u^\prime v^\prime}$, and $\widehat{u^\prime u^\prime}$ show 
progressively weaker response (in that order), since they are only influenced indirectly through turbulent transport.
As the artificial jet decays, the profiles for all components overlap with those of the planar shear layer.
Overall, these results indicate that  
while the flow components along $\boldvec{m}$ and $\boldvec{n}$ are not independent of one another (as expected in turbulent flows), 
the influence of $\widehat{w}$ on the shear layer is small and diminishes rather quickly.

\section{Summary and conclusions}
This study uses temporally evolving large eddy simulations to investigate three-component (or skewed) flow effects in turbulent shear layers. 
The main hypothesis of this study, encapsulated in figure~\ref{flowRegimesSchematic}, 
is that a skewed shear layer should evolve from an initial state that is best viewed in a coordinate frame aligned with the trailing edge of the splitter plate to a long-time state that should be viewed in a coordinate frame aligned with the imposed mean shear, 
with rotated mean velocity components $\widehat{u}$ and $\widehat{w}$ (aligned with and normal to the imposed mean shear, respectively).
In this view of the problem, any skewed shear layer should eventually evolve into a standard, canonical, two-component planar shear layer in that rotated coordinate frame, and the effects of skew should be intrinsically transient.
More specifically, we hypothesize that there should be an intermediate regime (``middle times'' here) where $\widehat{u}$ has become approximately self-similar and equal to a standard planar shear layer while $\widehat{w}$ remains finite, resembling a planar jet (since it must approach zero at the shear layer edges).

The results confirm this hypothesis. The simulations never reach the late times state, but the evolution clearly follows the hypothesized path.
For the particular initial conditions used here, 
the ``rotated spanwise'' component $\widehat{w}$ starts evolving like a decaying, approximately self-similar, planar jet from about $t^* \approx 5$, at which point its centerline velocity $\widehat{w}_c \approx 0.5 \Delta U$.
The component in the direction of the mean shear $\widehat{u}$ is within a few percent of its approximately self-similar profile from about $t^* \approx 10$, at which point $\widehat{w}_c \approx 0.35$.
Thus, the skewed shear layer obeys an approximate ``independence principle'' where the two flow components behave almost independently of each other, despite both being of similar magnitudes.
This approximate independence between the two flow components in the rotated frame is confirmed through a numerical experiment where an artificial spanwise jet was added to the planar shear layer state at one instant, resulting in rather small changes to the solution in the mean shear direction.

Another numerical experiment was conducted where the turbulence in the initial conditions were swapped between the planar and skewed cases.
These fictitious amalgamated cases were then found to closely follow the real case with the same initial mean flow, implying that the initial lack of alignment in the turbulence structures is of lower importance. This is an interesting finding since it differs from the main hypothesis in skewed boundary layers, where the misalignment of boundary layer turbulence structures is thought to be the major mediating factor~\citep{Lohmann_1976_ASME,Bradshaw_JFM_1985,Kiesow_JFM_2003,Lozano_JFM_2020}.
This difference between skewed boundary and shear layers is most likely due to the fact that the dominant turbulence structures are very different in the two.

Three-component effects are also observed in the organization of the turbulence. The misalignment in freestreams generates coherent structures
at the interface that are initially misaligned with the direction orthogonal to the mean shear.
These structures then undergo transient rotation to approach an alignment consistent with the rotated coordinate frame.
The coherence in the turbulence is found to be reduced during the transient evolution of a skewed shear layer.

\appendix
\section{Validation of boundary layer simulations}\label{appendixBL}
\begin{figure}
  \centering
  \includegraphics[width=0.6\textwidth]{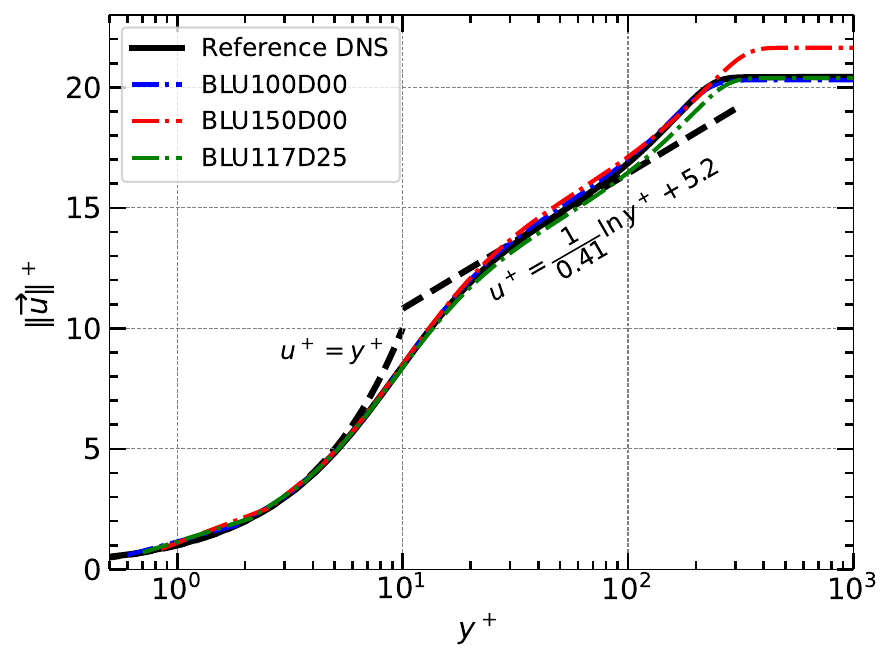}
  \caption{Wall-normal profiles of velocity magnitude for the boundary layer simulations (colored lines) in inner scaling, compared
  against the reference DNS data of \citet{Schlatter_JFM_2010} (black solid line). In addition, reference lines corresponding to the
  viscous sublayer and the log-layer are also overlaid as black dashed lines. See Table~\ref{tableTemporalBLcases} for reference to legend entries.}
\label{blUplusPlot}
\end{figure}

\begin{figure}
  \centering

  \begin{subfigure}[t]{0.45\textwidth}
      \centering
      \includegraphics[width=\linewidth]{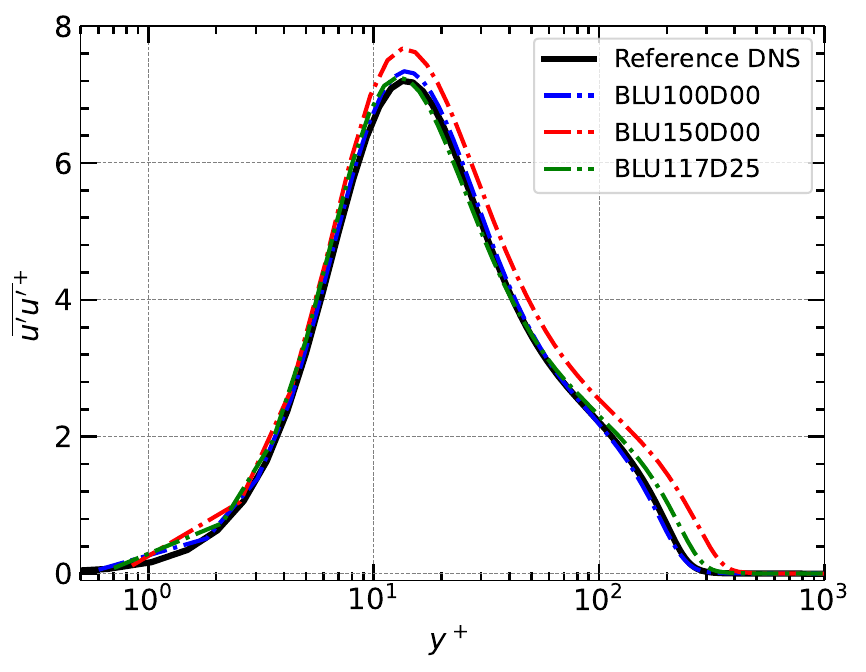}
  \end{subfigure}
  \hfill
  \begin{subfigure}[t]{0.45\textwidth}
      \centering
      \includegraphics[width=\linewidth]{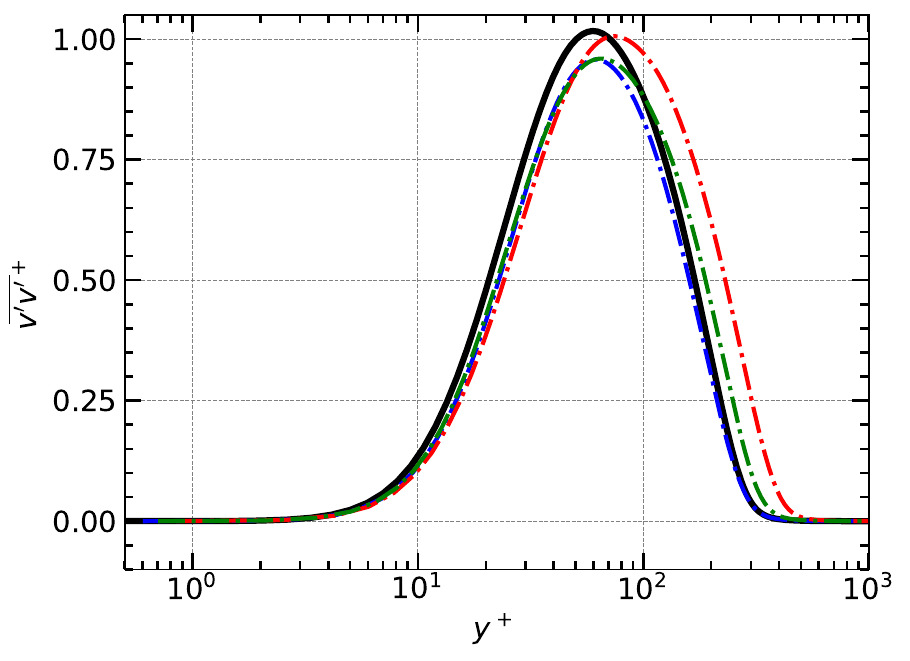}
  \end{subfigure}

  \begin{subfigure}[t]{0.45\textwidth}
      \centering
      \includegraphics[width=\linewidth]{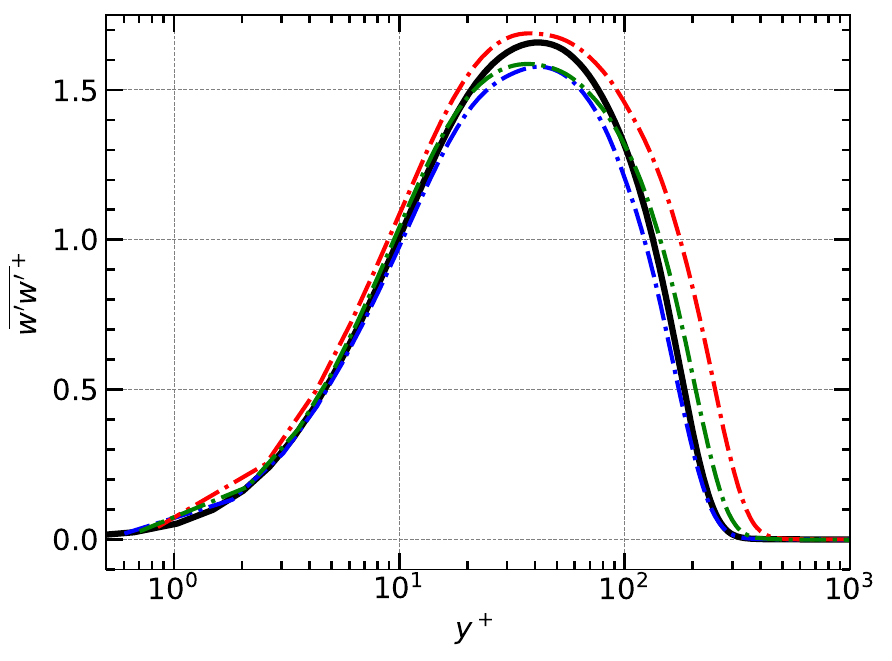}
  \end{subfigure}
  \hfill
  \begin{subfigure}[t]{0.45\textwidth}
      \centering
      \includegraphics[width=\linewidth]{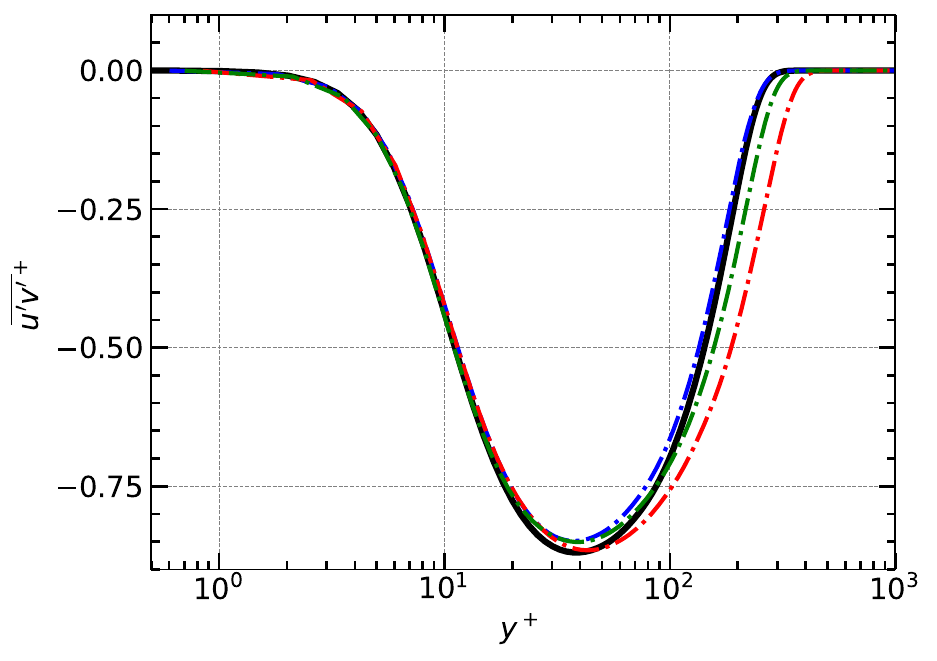}
  \end{subfigure}

  \caption{Wall-normal profiles of Reynolds stress tensor components for the boundary layer simulations (colored lines) in inner scaling, compared
  against the reference DNS data of \citet{Schlatter_JFM_2010} (black solid lines). See Table~\ref{tableTemporalBLcases} for reference to legend entries.}
  \label{blReStressPlot}
\end{figure}

The precursor boundary layer simulations are validated by
comparing the profiles of velocity magnitude 
$\|\boldvec{u}\|^+(y^+)$ and the non-zero Reynolds stresses 
$\widebar{u^\prime_i u^\prime_j}^+$ (superscript `+' denotes 
boundary layer inner scaling) against the DNS data of \citet{Schlatter_JFM_2010}
($Re_\tau \approx 252$) in Figs.~\ref{blUplusPlot} and \ref{blReStressPlot}, 
respectively. Note that since $\boldvec{U}_\text{H} \nparallel \boldvec{x}$
for boundary layer BLU117D25, its Reynolds stress tensor components are rotated by 
$-\theta_\text{H} (= -25^\circ)$ prior to comparison with other boundary layer cases. Overall,
both the velocity and Reynolds stress profiles for all three boundary layer cases are in good agreement with the DNS results, 
with some differences noted for the BLU150D00 case, possibly due to its relatively higher $Re_\tau$ (see Table~\ref{tableTemporalBLcases}) 
as compared to the DNS data.

\backsection[Supplementary data]{Movies showing the temporal development of pressure rollers in planar and skewed shear layers are available online. 
These visualizations complement the results shown in figure~\ref{combinedPressureRollers}.}

\backsection[Acknowledgements]{We acknowledge the access to HPC 
resources provided by the Argonne Leadership Computing Facility (ALCF),
Argonne National Lab and the Zaratan cluster at the University of Maryland.
Vedant Kumar also acknowledges the fruitful discussions with Md. Raihan Ali Khan
that have shaped the discussions in the manuscript.}

\backsection[Funding]{This work was supported by the Office of Naval Research (Grant ONR N00014-22-1-2038).}

\backsection[Declaration of interests]{The authors report no conflict of interest.}

\backsection[Data availability statement]{Data can be made available upon reasonable request.}

\backsection[Author ORCIDs]{Vedant Kumar (0000-0001-8502-3639), Dipendra Gupta (0000-0002-9889-8490), 
Gregory Paul Bewley (0000-0003-3421-3883), Johan Larsson (0000-0001-8387-1933)}

\bibliographystyle{jfm}
\bibliography{references}

\end{document}